\documentclass[%
 reprint,
nofootinbib,
 amsmath,amssymb,
 aps,
]{revtex4}

\begin{document}

\UseRawInputEncoding
\DeclareFixedFont{\CodelineFont}{\encodingdefault}{\familydefault}{\seriesdefault}{\shapedefault}{18pt}
\DeclareFixedFont{\fourteenpt}{\encodingdefault}{\familydefault}{\seriesdefault}{\shapedefault}{10pt}
 
%\preprint{APS/123-QED}

\title{Local thermodynamical equilibrium and relativistic dissipation} % Force line breaks with \\

\author{J. F\'elix Salazar}
\email{jose.salazar@correo.nucleares.unam.mx}
%\altaffiliation[]{}
\affiliation{Instituto de Ciencias Nucleares, Universidad Nacional Aut\'onoma de M\'exico\\
AP 70543, Ciudad de M\'exico 04510, M\'exico}
\author{Thomas Zannias}
\email{thomas.zannias@umich.mx}
\affiliation{Instituto de F\'\i sica y Matem\'aticas,
Universidad Michoacana de San Nicol\'as de Hidalgo,\\
Edificio C-3, Ciudad Universitaria, 58040 Morelia, Michoac\'an, M\'exico.}

%\author{J F\'elix Salazar}
%\altaffiliation[]{}
%%{jfelixsalazar@ifm.umich.mx}%Lines break automatically or can be forced with \\
%\email{jose.salazar@correo.nucleares.unam.mx}
%\author{Thomas Zannias}%
% \email{zannias@ifm.umich.mx}
%\affiliation{%
%Instituto de Ciencias Nucleares, Universidad Nacional Aut\'onoma de M\'exico\\
%AP 70543, Ciudad de M\'exico 04510, M\'exico\\
%Instituto de F\'isica y Matem\'aticas, Universidad Michoacana de San Nicol\'as de Hidalgo\\
%\normalsize{Edificio C-3 , Ciudad Universitaria, 58040 Morelia, Michoac\'an, M\'exico.}
%% This line break forced with \textbackslash\textbackslash
%}%


%\author{Thomas}
%\affiliation{
%\affiliation{
%\affiliation{%

\date{\today}% It is always \today, today,\\

\begin{abstract}
We introduce a class of relativistic fluid states satisfying  the
relativistic local thermodynamical equilibrium postulate (abbreviated as relativistic (LTE) postulate). 
States satisfying 
this postulate,
are states
 ``near equilibrium'' (a term defined precisely in the course of the paper)
and permit us to 
attach a fictitious ``local thermodynamical equilibrium'' state 
that fits event by event the actual fluid state. They single out an 
admissible class of rest frames
relative to which thermodynamical variables like the energy density, thermodynamical pressure, 
stresses, particle number density (or densities) measured by observers at rest relative
to these frames
are becoming frame independent provided
second (or higher) order deviations from the fictitious state of ''local 
thermodynamical equilibrium'' are ignored.
We have verified this property
for a large class  of  theories of relativistic dissipation that include
the Hiscock-Lindblom 
class  of first order theories,
the Eckart  and Landau-Lifshitz  theories, the Israel-Stewart transient thermodynamics, the 
Liu-M\"uller-Ruggeri theory, fluids of divergence type  and the latest developed  (BDNK)
theory. Moreover, 
the phenomenological
equations describing 
first order
deviations from the
fictitious ''local thermodynamical equilibrium'' state satisfy 
equations that
remain form invariant 
under change of frame within the
class of admissible frames.
We proved this property
for the Hiscock-Lindblom 
class  of first order theories
the Eckart and Landau-Lifshitz theories, 
the Israel-Stewart transient thermodynamics
and the 
Liu-M\"uller-Ruggeri theory of relativistic dissipation
the (BDNK) theory and we expect that the same property to hold for  the 
class of relativistic fluids of divergence type.

%The manifest invariance of the phenomenological equations
%under admissible frame is verified within the 
%transient thermodynamics while the same property has proven
%for  Liu-M\"uller-Ruggeri theory of relativistic dissipation. We also
%expect that states near equilibrium
%satisfying the (LTE) postulate to satisfy equivalent set of phenomenological equations
%although this property  remain to be verified.

\end{abstract}

\maketitle
\noindent
%\tableofcontents

\section{Introduction}

{\fourteenpt

We begin this paper, by  recalling the essential features of the first theory 
of Classical Irreversible Thermodynamics
 (CIT),
developed  by 
 Onsager \cite{OS1, OS2} and  independently by Eckart \cite{EC1, EC2} in  the decade of $1930s$-$1940s$.
 The theory deals with states of continuous Newtonian media
 that obey (or are compatible with) the 
  Local Thermodynamical Equilibrium
  (LTE) postulate. 
  This postulate primary provides a recipe that assigns an entropy to 
  off-equilibrium states and 
  presupposes  that 
  thermodynamical  equilibrium holds but only locally. The latter
 means that for any
 short time interval of the Newtonian time $t$ 
 and for any point within the medium, there exists a subsystem of volume\footnote{This volume $V$ is considered to be macroscopically small but microscopically large.} $V$
so that within this $V$  thermodynamical equilibrium prevails.
Evidence supporting the existence of such states
comes from the diversity that 
 continuous media
appear in nature. For instance, 
whenever states admit an inter-particle collision time scale
and this scale is much shorter
than any time scale that determines the evolution of
 macroscopic parameters,
then 
space time inhomogeneities 
in the macroscopic 
variables 
are ironed out by collisions
so that locally the state appears  to be uniform implying 
that (locally) thermodynamical equilibrium holds.\\

 The hypothesis 
  that within each subsystem $V$
 thermodynamical equilibrium prevails
 has important consequences. It asserts
the existence of
 local extensive variables  $(X_{1}, X_{2},...,X_{n})$ 
and a macroscopic entropy 
  $S=S(X_{1}, X_{2},...,X_{n})$
having the same 
functional form
as  the fundamental equation of 
  state describing states in global thermodynamical equilibrium.
 The additivity property of 
$S(X_{1}, X_{2},...,X_{n})$ implies that  
one can define an entropy density $s(x_{1}, x_{2},...,x_{n})$ 
function of 
$x_{i}=\frac {X_{i}}{V}, i \in (1,2,....n)$
and this
$s(x_{1}, x_{2},...,x_{n})$ 
plays a key role 
within the 
Onsager-Eckart theory of (CIT).
The (LTE) postulate asserts that
this 
$s(x_{1}, x_{2},...,x_{n})$ 
represents the physical entropy density of the
 underlying off-equilibrium state and this property  
combined with a set of balance laws, constitutive relations
 and by imposing the second law, closes the system of equations 
 for these 
 off equilibrium states.\\

To get more insights into the implications of the (LTE) postulate
and for later needs of the paper,
we take 
the underlying medium to be a collection of (classical) electrically neutral particles,
so that within each subsystem of volume $V$
 labelled by $(t, \vec x)$,
an inter-particle collision time scale
$\tau_{C}(t,  \vec x)$ is defined
obeying $\tau_{C}(t,\vec x)<<\tau_{M}(t,\vec x)$
where 
$\tau_{M}(t, \vec x)$
is any macroscopic  time scale
where the system change appreciably. 
The inequality
$\tau_{C}(t,\vec x)<<\tau_{M}(t,\vec x)$
asserts that 
within each $V$,
local thermodynamical equilibrium prevails
and
identifies the regime where the framework of fluid dynamics \cite{LLF} is applicable\footnote{
Systems 
obeying 
$\tau_{C}(t,\vec x)>>\tau_{M}(t,\vec x)$
are described by the one particle distribution function
and in that case one enters into the province of the kinetic theory of gases (for an introduction see for 
instance  \cite{Huang}). We also take the opportunity to clarify the difference between the 
term local thermodynamical equilibrium and the (LTE)-postulate employed  within (CIT).
Validity of the first simply requires
$\tau_{C}(t,\vec x)<<\tau_{M}(t,\vec x)$ while validity of 
 (LTE)-postulate at first presupposes 
 $\tau_{C}(t,\vec x)<<\tau_{M}(t,\vec x)$ 
 but most importantly, the postulate 
assigns
the non equilibrium entropy
to the underlying state.}. 
In that regime,
the subsystems
of volume $V$ are the familiar
 fluid elements or fluid cells
 and they are
 labeled  by a continuous real variables $\vec x=(x^{1}, x^{2}, x^{3})$
so that 
$(t, \vec x)$
labels  the fluid element at time $t$ centered around $\vec x$.
If $K$ is an inertial frame 
chosen so that the $(t, \vec x)$ cell is momentarily at rest, then 
relative to this frame
one can introduce (local) extensive variables
such as the internal energy $U(t, \vec x)$, the total particle number  $N(t, \vec x)$ and 
an additive entropy $S(t, \vec x)$ function
of $(U(t, \vec x), V, N(t, x))$.
Via these variables, the (LTE) postulate assigns a physical entropy 
to the underlying state and as a consequence of this 
property, these variables 
satisfy a host of
thermodynamical relations that we now briefly mention.\\
For their derivation, let us for the moment consider 
a (global) equilibrium fluid state
at rest relative to a global inertial frame, 
and denote by
$(S, U, V, N)$
the 
spatially and temporally 
homogenous 
total entropy $S$, internal energy $U$, and particle number $N$ within
a macroscopic volume $V$. These variables,
 under
 an infinitesimal reversible transformation,  
satisfy the Gibbs equilibrium relation 
\begin{equation}
dS=\frac {1}{T}dU+\frac {P}{T}dV-\frac {\mu}{T} dN
\label{GIBBS}
 \end{equation}
where 
$(T, P, {\mu})$
are the temperature $T$, pressure $P$ and chemical potential $\mu$ of
the particle species.
The homogeneity property of $S(U, V, N)$ under rescaling implies
that  
\begin{equation}
S(\lambda U, \lambda V, \lambda N)=\lambda S(U, V, N),\quad \lambda >0
 \label{SCAL}
 \end{equation}
and by choosing $\lambda=1+\epsilon$ and expanding
 the identity 
  $$S(U+\epsilon U, V+\epsilon V, N+\epsilon N)=(1+\epsilon) S(U, V, N)$$
then one gets the fundamental relation:
\begin{equation}
U=TS-PV+\mu N,
\label{SP}
 \end{equation}
which implies the Gibbs-Duhem relation:
\begin{equation}
SdT-VdP+Nd\mu=0.
\label{GibbsDD}
\end{equation}  
  
  Dividing
(\ref{SP}) by $V$ and introducing the internal energy density $e$, entropy density $s$ and particle density $n$, 
then 
(\ref{SP})
yields
\begin{equation}
e=Ts-P+\mu n
\label{Chem}
\end{equation}
which via differentiation and in  combination with the  densitized form of the Gibbs-Duhem relation (\ref{GibbsDD}) 
i.e. $sdT-dP+nd\mu=0$,
implies
\begin{equation}
de=Tds +\mu dn.
\label{Cibbs22}
\end{equation}
Thus knowledge of the equation of state $e=e(s, n)$ determines
via differentiation the equilibrium temperature and  chemical potential $\mu$ of the underlying equilibrium fluid state.\\ 
Formulas
(\ref{GIBBS}-\ref{Cibbs22}) are the well known
 relations that fluid states in global thermodynamical equilibrium must satisfy
and we derive them here just to illustrate 
the implications that the (LTE) postulate has upon the description of hydrodynamical states.
Whenever
a state is compatible with this postulate,,
then
$U(t, \vec x), N(t, \vec x), S(t, \vec x)$ 
and the corresponding densities $(e(t, \vec x), n(t, \vec x), s(t, \vec x))$ 
satisfy relations
  (\ref{GIBBS}-\ref{Cibbs22})
 i.e. the same relations as if the state was in global thermodynamical equilibrium.
 However a major difference should be noted:
the variables  $(U(t, \vec x), N(t, \vec x), s(t, \vec x), e(t, \vec x), n(t, \vec x), s(t, \vec x))$
 obey 
 (\ref{GIBBS}-\ref{Cibbs22}) only locally i.e. at 
 the interior of (any) fluid cell. \\    

The (CIT) of Onsager-Eckart has been proven to be a successful theory.
 It leads to the Fourier-Navier-Stokes fluid system
 which is
 the standard theory 
that describes terrestrial and a large class of astrophysical fluid
flows and the classical book by de Groot and Mazur \cite{dGM}
deals extensively with the development and applications of 
the 
Onsager-Eckart theory of 
(CIT) to linear thermodynamics of irreversible processes
(the interest reader is refereed to that book for further discussion and additional references).\\
However, despite these successes,
 years of efforts and experience lead to the realization that  
 not all goes that well with the dynamics of states compatible with 
  the (LTE) postulate within the (CIT). 
The  postulate dictates a very rigid 
functional dependance of the physical entropy density 
 $s$ upon the local thermodynamical variables,
 and this
 leads to the prediction
 that 
 disturbances in temperature and stresses
 propagate with an unbounded speed,
 a highly unexpected and counterintuitive prediction.\\
 M\"uller \citep{Mul1} in a fundamental paper written in $(1967)$, suggested 
 a way out of this problematic issue. He proposed that
 the physical entropy density $s$ of non equilibrium fluid states
should 
 receive contributions from
dissipative variables
 such as heat flux, bulk and shear stresses,
a proposition which is in a blatant contradiction to the 
spirit of the (LTE) postulate.\\
M\"uller $(1967)$ suggestion 
lead to the development of extended theories
of irreversible thermodynamics and pave the way 
to the formulation of 
 the entropy principle introduced by Coleman and Noll \cite{ColN1, ColN2}, M\"uller \cite{MulN1, MulN2} 
 and others, 
 as well as lead to the realization that
 systems of symmetric-hyperbolic equations 
should be the appropriate set of dynamical equations describing irreversible thermodynamics (for a historical development, references and current status of 
extended theories, consult refs \cite {JVL, Mul4, Mul7}).\\
Although
the developments of extended theories
remove the assertion of the (LTE) postulate 
that dictates the 
dependance of
the physical entropy density $s$ 
 upon the local thermodynamical variables,
 nevertheless
 in these extended theories 
the validity of the hydrodynamical regime is left intact 
and thus also the validity 
of local thermodynamical equilibrium is preserved.  This in turn implies
the existence of
(local) hydrodynamical variables $(X_{1}, X_{2},...,X_{n})$
associated with the underlying fluid state.
Using these variables,
one may
still employ 
$s(x_{1}, x_{2},...,x_{n})$
as an
''equilibrium equation of state'' 
and via the  Gibbs formalism,
attaches a ''fictitious\footnote{
We refer to this state
as a 
 ''fictitious'' state
because the entropy that is defined
by 
  $s(x_{1}, x_{2},...,x_{n})$ bears
no relation to the physical entropy of the underlying state.}
''local thermodynamical equilibrium" state
to the underlying fluid state.
This 
fictitious ''local thermodynamical equilibrium state''
it is not always of physical relevance but nevertheless
it can be  very convenient\footnote{Traces of this  fictitious ''local thermodynamical equilibrium state'' can be seen in the introductory treatments of extended thermodynamics
(see for example
\cite{JVL}, \cite{RZ}),
where one often encounters terms
 like the equilibrium temperature, pressure, etc. These equilibrium quantities
 are generated by employing  
  the equilibrium equation of state
  $s(x_{1}, x_{2},...,x_{n})$ alluded above.}.
As we shall see further ahead, upon passing to the relativistic description,  the idea of attaching 
a fictitious local ''thermodynamical equilibrium state'' to a fluid state, will be proven essential  in the development of this work.\\
The present paper discuss an adaptation of the (LTE) postulate 
to the 
 relativistic regime 
 so that the postulate becomes a tool for the analysis of
relativistic fluids states. 
The relativistic framework offers the possibility  
to introduce a special class of relativistic fluid states 
refereed as states compatible with the relativistic 
(LTE) postulate and their precise definition will be introduced in the following sections.
Here,
we offer a few comments regarding the 
motivations that lead us to introduce this special class of states.\\
For this, it is of relevance to 
recall the observational breakthroughs 
that took place 
within the last ten years or so, confirming the
minute predictions of Einstein's general relativity or of relativistic physics.
It is sufficient
to recall the monumental detection of 
  gravitational waves by the LIGO observatory,  
  the  binary neutron star 
 inspiral and merger 
 GW170817 detected by the LIGO/VIRGO gravitational wave observatory network,
 the photographs
released by 
 the 
 Event Horizon Telescope (EHT) collaboration 
 of the event horizon'' of the supermassive black hole at
 the center of the $M87$ and at the center of our galaxy,
the ongoing experiments
at the  
Relativistic Heavy Ion Collision at BNL and at Large Hadron Collider at CERN,
and the 
formation of the 
terrestrial
``mini bing-bangs''
in the form of 
quark-gluon plasma.\\
Although presently, the scientific community is in the process of digesting these 
observational breakthroughs, nevertheless one
message comes across clearly: relativists and high energy physicists
alike need to develop
reliable theories
of relativistic continuous media in order to confront the new observational realities.
It has been however realized and in fact long long ago,
 that relativistic dissipation is a tough mathematical 
problem which is still wide open
(refs.\cite{Nor2},\cite{RZ}, \cite{RR}
offers un update on the latest developments). As we shall discuss further ahead, all so far proposed theories 
 are highly non linear and they are plagued by 
 conceptual issues such as causality violation \cite{His2}, instability of equilibrium states
 (for a recent discussion see
 \cite{GAV2} \cite{GAV3}, \cite{GAV6}) lack of a sensible definition of a fluid four velocity (see for instance 
 \cite{vKAM}, \cite{Isr1}, \cite{Isr2} and the next two sections of this paper) 
 and often the dynamical evolution equations
  fail to constitute a 
 symmetric-hyperbolic and causal set of equations (see for example \cite{Mul6}, \cite{Ger1}).\\ 
 In this rough territory, progress has accomplished
 by restricting attention to equilibrium states and 
 the analysis of their perturbations.
The work in 
\cite{His2} studies the stability properties of first order theories, while in 
refs. \cite{His2},\cite{Olson} the stability properties of transient thermodynamics has been addressed.
The stability of equilibrium states and causality properties of Carter's theory of heat conducting relativistic fluids 
(for an introduction to this theory see \cite{Car1} \cite{Car2} \cite{Car3})
has been the focus of refs \cite{HisO},\cite{GAV13},\cite{GAV7}.
%while the second one employs Gibbs maximum principle techniques as elaborated in 
%$\cite{GAV7}.
 \\

Even though  this brief survey, shows a sizable amount of scientific activity 
on equilibrium states and their properties, still others aspects of relativistic dissipation needed to 
be investigated. The aim of the present work is to introduce a family of states 
as an alternative to
 equilibrium states.
 This class has been motivated by the properties of states satisfying the (LTE) postulate within the (CIT)
and they will be referred as states 
satisfying the  
relativistic  (LTE) postulate. 
They are states that on the one
 hand, go beyond the class of the equilibrium states and 
 on the other hand, 
 avoid (partially) the complexities associated with the non linear nature
 of the underlying fluid equations. The central idea 
 that underlies 
 this special class of states 
 is their
property that they are 
 states ''near equilibrium''
a term introduced by Israel long time ago
 and used as the building block in the development of  transient thermodynamics.
 Although the 
 term states ''near equilibrium'' at a first side suggest that such states could be just perturbations of equilibrium states, 
  that is not the case and that will become clear in the course of the paper.
The identification of states 
satisfying the  
relativistic  (LTE) postulate 
is almost independent upon the 
underlying fluid theory. Their definition requires only that 
within the the underlying theory, states  to be described by a conserved symmetric energy 
 momentum tensor
 and a collection of conserved particle  currents (this last requirement is optional). \\
The key ingredient that allows us to introduce this special class
of states is the possibility to attach a 
fictitious ''local thermodynamical equilibrium'' state
to an underlying fluid state
and this attachment proceeds along the same reasonig 
as for the Newtonian
framework. However, in the relativistic domain
local observers and their measurements 
in combination to
an ''equilibrium equation of state'' 
are the key elements
that  lead to this
fictitious ''local thermodynamical equilibrium'' state.
This fictitious state in combination 
with 
states near thermodynamical equilibrium (the precise coordinate free definition of such states
is discussed in latter sections)
yield the class of fluid states  compatible with the relativistic (LTE) postulate and details of 
their construction are discussed in the following sections.

\section{ On dissipative relativistic fluid theories}

In this section, 
we prepare the ground for 
 the formulation of the relativistic version of the (LTE) postulate
 and for this purpose,
 at first, we discuss briefly the salient features of 
 a few theories 
 of dissipative relativistic fluids 
developed so far.\\
 
 Early attempts to construct 
viable theories initiated with the work of Eckart \citep{Eck}
 in the $(1940)s$,
 the work 
 by Landau and Lifshitz \cite{LaL} in the $(1950)s$ and culminated in the formulation of the class of first order theories 
developed  by Hiscock and Lindblom \cite{His2} in the $(1980)$s.
 At around $1970$s, 
 Israel \cite{Isr1} and Israel and Stewart \citep{Isr2}
formulated the transient thermodynamics 
 which is a second order theory
 a term that will become clear further ahead.
Within these theories,
states of a relativistic fluid
  are described by a symmetric energy momentum  tensor $T^{\mu\nu}$ a particle  current\footnote{For simplicity, in this work we treat the case of a simple fluid. For a fluid mixture, one may introduce $n$ conserved particle currents  $J_{i}^{\mu}$, $i \in (1,2,...n)$ and proceeds in a similar manner. } $J^{\mu}$
obeying the conservation laws
\begin{equation}
\nabla_\mu T^{\mu \nu} = \nabla_\mu J^\mu = 0,
\label{BE}
\end{equation}
and accompanied by an entropy flux four vector $S^{\alpha}$ obeying
\begin{equation}
\nabla_{\alpha}S^{\alpha}=\sigma,
\label{EL}
\end{equation}
where $\sigma \geq 0$ is the entropy production scalar.\\
With the development of Rational Extended Irreversible Thermodynamics 
in the late $1980s$, by 
M\"uller, Liu, Ruggeri and collaborators,
a new trend in the field of relativistic dissipative fluids opened up
(for and introduction consult ref.\cite{Mul4,Mul7}). 
 New theories
 have been developed whose states are described by
  a collection of tensor fields
  obeying a manifestly symmetric-hyperbolic (or causal)
systems of dynamical equations.
Such  theories include
the theory developed by
 Liu, M\"uller and Ruggeri
 (see \cite{Mul6,Mul4}), the theory of relativistic fluids of divergence type
developed by Pennisi \cite{Pen}, 
and independently by Geroch and Lindblom
\cite{Ger1}, Carter's 
theory of 
heat conducting fluids \cite{Car1, Car2, Car3},
the Unified Extended Irreversible Thermodynamics (UEIT) developed by 
Gavasinno and Antonelli \cite{GAV7} ,\cite{GAV1}. 
%, and further references therein \cite{Ger1}, \cite{HisO}, \cite{GAV13}).
\\
Finally, we should mention 
the enormous impact
the recent experiments on heavy ion collisions 
had 
 upon the development of theories of relativistic dissipation
 (for details on this connection
consult for instance \cite{Hei}). Here the unexpected realization that models of relativistic viscous hydrodynamics describe reliably observational dada
acted as a stimulus for the development of new theories.
Such theories include
 the 
Denicol-Niemi-Molnar-Rischke (DNMR) theory \cite{DNMR},
the  Baier-Romatschke-Son-Starinets-Stephanov
theory \cite{BRSSS},
or  anisotropic hydrodynamics theories (for an introduction see see \cite{AH1}, \cite{AH2}) etc.
These theories are second order theories derived either
as limits of the relativistic Boltzmann equation case of (DNMR) theory, or 
as  effective theories obtained from the gradient expansion within the quantum field theory, truncated at second order, case of 
Baier-Romatschke-Son-Starinets-Stephanov (BRSSS) theory.\\
The quest to 
develop theories with physically relevant characteristics, for instance 
theories that their equilibrium states are stable, respect causality, the corresponding dynamical equations admit a well posed initial value formulation,
 lead to the development of the 
 (BDNK) formalism 
 (for an introduction see
 (\cite{Kov1}, \cite{Kov2}, 
\cite{Nor1}, \cite{Nor2} ) 
which is a formalism that lead to the development of first order theories with
 remarkable properties
 and in section $(IV)$ we shall have the opportunity to comment more on this formalism.\\
 
  It is not the purpose of this paper to provide a detailed description of 
the above mentioned theories neither to discuss 
their virtues and failures (for recent reviews see for instance \cite{Nor2}, \cite{Mul7}, \cite{RMa1}, \cite{FT1},
\cite{FLO},  \cite{RR}, \cite{RZ}, \cite{Isr5}). 
Our purpose
is to
adapt the
(LTE) postulate 
from its natural habitat i.e. the regime of Newtonian continuous 
media, to the
realm of relativistic fluids
and our approach
has been motivated by the development of the Israel-Stewart transient thermodynamics
although
the framework that we shall develop is
 applicable to arbitrary theories
of relativistic dissipation\footnote{The only restriction we shall impose 
on the structure of the
underlying theory 
is that  amongst the variables that describe fluid states, includes a conserved particle current
  $J^{\mu}$ and a symmetric energy momentum tensor $T^{\mu\nu}$ which satisfies the weak energy condition.}.\\
As we have already mentioned, the description of relativistic dissipation presents a few conceptual challanges 
that  are
absent, for instance,
 from states of the Newtonian 
Fourier-Navier-Stokes system.
For the later, the 
 field equations are balance laws expressing conservation of mass, energy and linear momentum
and  involve
  the velocity field, 
 mass density,  the components of a stress and the components of the heat flux.
 By invoking the entropy principle, constitutive relations 
 are proposed so that
 one gets a closed system of equations for the components of
 the velocity field, energy density and temperature, all of them defined relative to a fixed global inertial frame.
 %In this effective system,  the fluid's velocity field and temperature determine the components of the bulk, shear  and heat flux respectively.
 \\
   Unfortunately, attempts to interpret dissipative states of a relativistic fluid in a similar manner
 run into difficulties. Terms like 
  particle and energy densities, stresses, temperature etc. are in general observer dependent quantities
    and more dramatic is the realization that 
   the fluid's four velocity
is either non existent or if it is defined 
looses its prominence\footnote{An exception 
  is provided by states describing a simple perfect fluid. Here a four velocity field 
  $u^{\mu}$  is assigned dynamically by the theory and 
  perfect fluid states are described
  by 
  the future-directed, timelike velocity field $u^{\mu}$ and  
   two spacetime scalar fields $(n,\rho)$  representing the particle density and energy density
  measured by a comoving with the flow  observer. The particle number and energy momentum conservation laws
supplemented by an equation of state fix these field uniquely.}.
For fluid theories whose states are described 
partially\footnote{The use of the world ''partially'' reflects the fact that we have not specified the underlying fluid theory. It may happen that 
  for an arbitrary theory,
  one may need besides
  $(T^{\mu\nu}, J^{\mu})$   
additional variables to specify a fluid state for instance the entropy flux $S^{\mu}$ or other additional fields.} 
by a conserved particle current $J^{\mu}$ and a
 conserved energy momentum tensor
$ T^{\mu\nu}$, one has the liberty to define the fluid's four velocity as 
%the class of observers defined by the 
 the unique, future directed timelike eigenvector
 $ u_{E}^\mu $
 of the energy momentum tensor $T^{\mu\nu}$ i.e. identify the fluid's four velocity with the flow of energy\footnote{This $ u_{E}^\mu $ is well defined provided the energy momentum tensor obeys the weak energy condition.}
or one may identify the fluid four velocity
with the 
$4$-velocity
$ u_{N}^\mu $ defined 
via 
 $J^{\mu} = nu_{N}^{\mu}$
 where
 $n$ is the particle density measured by 
an observer comoving with  
 this $u^{\mu}_{N}$
 and thus identify the fluid four velocity with the particle motion.
 Should one identify 
 the fluid's four velocity 
with 
 $ u_{E}^\mu$,  $ u_{N}^\mu $ 
 or may one employ an altogether different four velocity field? This ambiguity
 regarding the 
 choice of the fluid's four velocity is 
 a common feature of all so far proposed theories of relativistic dissipation
 and to this day, there is not a satisfactory resolutions of this dilemma\footnote{
 Viewing relativistic fluids as arising
 from a microscopic descriptions either from relativistic kinetic theory or 
 from the expectation values of
 quantum observables via suitable averaging, it is hoped that this transitions may single our a unique fluid flow field. Unfortunately
 up to to now there is no such a satisfactory prediction.}.\\
It is however worth noticing
that although the notion of a fluid four velocity
may be ambiguous or even ill defined, 
the general relativistic framework
allows to introduce a smooth, timelike future directed ''velocity field''
$u^{\mu}$
 in the spacetime region occupied by the fluid and 
view
this $u^{\mu}$
as providing  a field of ``rest frames'' employed by the $u^{\mu}$-observers 
who probe the state of the fluid.
This liberty in the choice of 
a potential ``velocity field'' $u^{\mu}$,
combined with a particular class of relativistic fluid states
defined below,
are the building blocks
for the formulation of states satisfying the relativistic (LTE) postulate  
and in the next section we address this construction.\\

\section{ On states  compatible with the relativistic (LTE)-postulate}

In this  section, we consider a test relativistic fluid propagating on a smooth background spacetime $(M,g)$
and
within the region occupied by the fluid, we introduce a 
smooth, non singular, future pointing velocity field  
 $u^\mu$ whose 
 sole purpose
 is to provide
a family of local observes who perform measurements upon the fluid's state.
At any event $p$ in the fluids interior 
and relative to the
  local rest frame
whose time axis at $p$ coincides with
$u^\mu(p)$,
we 
consider an infinitesimal element of 
a spacelike $3$-volume $V$, orthogonal a $u^\alpha(p)$
and view this $V$ as the local fluid cell.
Furthermore, we assume that it is possible to choose this
 $u^\mu$ 
so that 
relative
to the rest frame defined by
 $u^\alpha(p)$, a collision time scale 
$\tau_{C}(p)$ is defined
that 
satisfies $\tau_{C}(p)<<\tau_{M}(p)$
where 
$\tau_{M}(p)$
is another time scale\footnote{
These time scales are identical to those that we defined earlier on for Newtonian fluids,
although 
there is however a pronounced difference between the Newtonian and relativistic
 case. While in the former case
 the inequality $\tau_{C}(p)<<\tau_{M}(p)$
 is frame independent, in the relativistic regime and due to time dilatation effects 
in general  this inequality is frame dependent.  
States compatible with the relativistic version of (LTE) postulate
 that we are about to introduce, is
 a class of states that 
 allows to introduce a preferable class of frames, refereed as admissible frames,
 having the property
 that if  $\tau_{C}(p)<<\tau_{M}(p)$
 holds in one frame within this admissible 
 class, then frame changes within this admissible class, leaves this inequality intact. }
determined by the velocity field, for instance
by its expansion, shear, rotation, etc.
As for the Newtonian case, this inequality implies that 
the $u^\alpha(p)$-observer 
will conclude that 
local thermodynamical equilibrium prevails within $V$
and
 using
$T^{\mu\nu}, J^{\mu}$ and $S^{\mu}$,
 assigns at $p$
a particle number $n(p)$, energy density $\rho(p)$
and an entropy density  $\hat s(p)$
via
\begin{equation}
n(p)=-J_{\mu}u^{\mu},\quad\quad \rho(p)=T_{\mu\nu}u^{\mu}u^{\nu},\quad\quad \hat s(p)=-S_{\mu}u^{\mu}.
\label{FUND}	
\end{equation}
Using this $u^{\mu}$ one 
in general 
%this $u^\alpha(p)$-observer will
decomposes\footnote{In 
(\ref{OD}-\ref{SD}), and in the sequel,
we
write 
$\rho(u), P(u), h(u), \tau(u)^{\mu}{}_{\nu}$ etc., in
order to remind the reader that these 
variables are 
measured by the $u-$observer. This notation further signifies 
that  the
decompositions  in 
(\ref{OD}-\ref{SD})
are frame dependent.
Often we write 
$n(p)$, $\rho(p)$ etc to denote results of measurements by the $u$-observer at a particular event $p$.
} 
$T^{\mu\nu}$ and $J^{\mu}$
according to:
\begin{equation}
T^{\mu\nu}=\rho(u)u^{\mu}u^{\nu}+P(u)\Delta(u) ^{\mu\nu}+h(u)^{\mu}u^{\nu}+h(u)^{\nu}u^{\mu}+\tau(u)^{\mu\nu},
\label{OD}
\end{equation}
\begin{equation}
	J^{\mu}=n(u)u^{\mu}+n(u)^{\mu},
\label{OD_11}	
\end{equation}
where
$h(u)^{\mu}, n(u)^{\mu},\tau(u)^{\mu\nu}$
stand for 
the energy flow vector, the particle ''drift'' 
and the spatial symmetric pressure tensor $\tau (u)^{\mu\nu}$ respectively all of them measured by the $u^{\mu}$ observer. In these expansions, the pressure tensor
$\tau (u)^{\mu\nu}$
defines 
the bulk pressure
$\pi(u)$
and shear stresses
$\pi(u)^{\mu\nu}$
according to 
\begin{equation}
\tau(u)^{\mu\nu}=\pi(u)\Delta(u) ^{\mu\nu}+\pi(u)^{\mu\nu},\quad \pi(u)^{\mu}_{\mu}=0,
\label{SD}
\end{equation}
the fields $(h^{\mu}, n^{\mu})$ and $\tau(u)^{\mu\nu}$ satisfy
\begin{equation}
h(u)^{\mu}u_{\mu}=n(u)^{\mu}u_{\mu}=u_{\mu}\tau(u)^{\mu\nu} = 0,
\label{OD_22}	
\end{equation}
while
$\Delta ^{\mu\nu}(u)=g^{\mu\nu}+u^{\mu}u^{\nu}$ stands for the projection tensor.\\
 Since as we have already mentioned, the inequality
$\tau_{C}(p)<<\tau_{M}(p)$,
implies validity of the local thermodynamical equilibrium within $V$, 
 by appealing 
to Gibbs formulation{\footnote{For an introduction to this formulation see for instance \cite{Cal}.} of equilibrium thermodynamics,
this $u^{\mu}$ observer
will conclude that
$n(p)$ y $\rho(p)$ must satisfy 
an "equilibrium equation of state'' of the form
$s=s(\rho, n)$. In turn this $s=s(\rho, n)$ implies
\begin{equation}
ds =\frac {1}{T} d\rho - \Theta d n,
\label{OD_22}	
\end{equation}
thus defining the local
 temperature $T(u)$ and the thermal potential $\Theta(u)$
as measured by the $u^{\mu}$-observer,
 while the fundamental relation $s(u)=(\rho(u)+P(u))T^{-1}(u)-\Theta(u) n(u)$ with $s(u):=
 s(\rho(u), n(u))$ defines the (local) equilibrium pressure
$P(u)$.
Therefore, once a velocity field
  $u^\mu$ and an equation of state $s=s(\rho, n)$ have been specified, then 
  $(T^{\mu\nu}, J^{\mu})$ define at $p$ a
 ''local thermodynamical
equilibrium'' state specified by

\begin{equation}
(n(p), \rho(p), P(p), s(p), T(p), \Theta(p)).
\label{23}
\end{equation}

As long as
the velocity field $u^{\mu}$ 
and the fluid state allow the inequality
$\tau_{C}(p)<<\tau_{M}(p)$
to hold at any event $p$ within the fluid's interior, 
then the variables in (\ref{23})
are defined over the entire spacetime region occupied by the fluid
and combined with $u^\mu$ give rise to a (local)
 particle current\footnote{It is understood that
 in
 (\ref{33}), the fields
$(n, \rho, P, s)$ stand for $(n(u), \rho(u), P(u), s(u(n(u), \rho(u))$.} 
  $J^{\mu}_{0}$, a  tensor $T^{\mu\nu}_{0}$ and an
 ''entropy flux'' vector $S^{\mu}_{0}$, 
via
\begin{equation}
J^{\mu}_{0}=nu^{\mu},\quad
T^{\mu\nu}_{0}=(\rho+P)u^{\mu}u^{\nu}+Pg^{\mu\nu},\quad S^{\mu}_{0}=s(\rho, n)u^{\mu}.
\label{33}
\end{equation}
\\
These fields, in the terminology of Israel-Stewart \cite{Isr1, Isr2}, define
a ``local equilibrium reference state'' attached to the physical state described by
the variables 
  $(T^{\mu\nu}, J^{\mu}, S^{\mu} )$.
This ''local equilibrium reference state''
depends upon the (almost arbitrarily) chosen 
velocity field $u^{\mu}$ and thus at a first sight seem to be of limited
utility.  Any other 
$\hat u^\alpha(p)$-observer
and as long as at $p$ local equilibrium prevails relative to 
%the rest 
her/his frame, %of this $\hat u^\alpha(p)$ observer 
will measure  
\begin{equation}
(\hat n(p), \hat \rho(p), \hat P(p), \hat s(p), \hat T(p), \hat \Theta(p))
\label{H22}
\end{equation}
which are
related 
to those in (\ref {23}) via complicated transformations
formulas induced by a local Lorentz transformation that relates
the frames defined by $u^\alpha(p)$ and $\hat u^\alpha(p)$. However, for a particular states, the fields 
 in (\ref{23}, \ref{H22})
transform under 
local Lorentz transformations in a simple manner 
so they  become useful tools in describing properties of the
%the physics of the 
underlying fluid state. 
\\
In order to identify these states, we consider again
the unique timelike eigenvector 
$ u^\mu_E $
of the energy momentum tensor 
 $T^{\mu\nu}$ which specifies
 the  energy frame, 
and the timelike and future directed vector
$ u^\mu_N$ that specifies the particle frame.
  For an arbitrary 
  fluid state determined (partially) by
  $(T^{\mu\nu}, J^{\mu})$   
  the
  fields 
  $( u^\mu_E ,u^\mu_N) $
are in general distinct and
 there is not any obvious relation between the two. However, 
  by appealing to an idea introduced by Israel long time ago\footnote{Israel suggested that for states near equilibrium, a theory can be developed 
 that describes the first order deviations from 
 a local ''equilibrium reference state''  
 and this theory  
  is invariant under particular change of rest frames within the class of admissible frames.
  This idea was elaborated further in a M.Sc. thesis written by Aitken \cite{Ait} then student of Israel.}
and employed in the formulation of the transient thermodynamics,
we introduce states referred as ''states near equilibrium'' (or ''states close to equilibrium'')
in the following manner.
 At an 
  event $p$ within the fluid's region, 
one sets up an orthonormal frame with a time axis parallel to
 $u_{E}$ accompanied by 
 a triad $e_{i},~i \in (1,2,3)$ of spacelike vectors
 so that 
 $(u_{E}, e_{i}),~i \in (1,2,3)$,  
 constitutes an orthonormal tetrad.
 Since
 there is freedom in the choice of the triad $e_{i},~i \in (1,2,3)$     , 
without loss of generality, one may assume:
 \begin{equation}
 \begin{split}
 u_{N}^{\mu} = & [1-\frac {v^{2}}{c^{2}}]^{-\frac {1}{2}}u_{E}^{\mu}+\frac {v}{c}[1-\frac {v^{2}}{c^{2}}]^{-\frac {1}{2}}e_{1}^{\mu} \\
 = & cosh\epsilon~u_{E}^{\mu}+sinh\epsilon~ e_{1}^{\mu},\quad g(u_{E}, e_{1})=0,
 \label{ELL}
 \end{split}
 \end{equation}
where $\vec v=ve_{1}$ is the ``relative velocity" of the particle
frame relative to the  
energy frame.
This relation defines a  pseudo angle 
$\epsilon$ between $u_{E}$ and $u_{N}$ 
according to:
\begin{equation}
	cosh \epsilon=-g(u_{E}, u_{N})=\left[1-\left(\frac {v^{2}}{c^2}\right)\right]^{-\frac {1}{2}},
\label{PSA}
\end{equation} 
and this angle  plays an important role.
It is suffice to mention that  for all  proposed theories
of relativistic dissipation,
it holds that for
states in a global equilibrium the fields 
$u_{E}$ and  $u_{N}$ coincide and thus $\epsilon=0$.\\
Motivated by this property 
and in the spirit of transient thermodynamics, 
we defines for a simple fluid
states near\footnote{For a fluid mixture consisting of $n-$particle currents $(J_{1},J_{2},....,J_{n})$, 
one may define $n$-four velocities $(u_{1},u_{2},....,u_{n})$
and thus introduce $n-$pseudo angles
$(\epsilon_{1},\epsilon_{2},....,\epsilon_{n})$ between $u_{E}$ and the corresponding
$(u_{1},u_{2},....,u_{n})$. A state then is close to equilibrium, whenever
$(\epsilon_{1},\epsilon_{2},....,\epsilon_{n})$ 
obey  $\epsilon_{i}\leq 1$ for all $i \in (1,2,.....,n)$.} 
equilibrium\footnote{Further below, we shall 
introduce their close relatives named states satisfying with the relativistic (LTE) postulate.}
as those  states 
that have the property that
the 
pseudo-angle $\epsilon$ 
in 
(\ref{PSA})
satisfies
everywhere within the region occupied by the fluid
the condition 
$\epsilon=\frac {v}{c} <<1$.
For such states, one notices
from  
 \begin{equation}
 J^{\mu}=n_{N}u_{N}^{\mu}=n_{N}(cosh\epsilon ~u_{E}^{\mu}+ sinh\epsilon~e_{1}^{\mu})=n_{E}u_{E}^{\mu}+ n^{\mu},
  \label{PELL}
 \end{equation}
and thus the densities $n_{N}$ and $n_{E}$ measured by the
 $ u^\mu_N $
respectively the $u^\mu_E$ observers  
 satisfy
\begin{equation}
n_{E}=n_{N}cosh\epsilon=n_{N}+O(\epsilon^{n}),\quad n\geq 2
\label{PELLL}
 \end{equation}
 implying that 
$n_{N}$ and $n_{E}$
are considered to be frame independent
provided terms of order $\epsilon^{2}<<1$ and higher are neglected\footnote { Here after we follow the notation
 of ref.\cite{Isr1} often we set 
 $\epsilon :=O_{1}$ while terms like $O_{2}, O_{3}.....$ 
 signify terms of first, second, third order... deviations.}.  
 This (approximately) 
``invariance property''
of the particle density
holds also for other fields 
that appear in (\ref{23}) and (\ref{H22}) and 
in order to investigate in a systematic manner  
the transformation properties 
of the thermodynamical variables under frame change, 
let $(u^{\mu}, \hat u^{\mu})$ 
be two (future pointing) unit timelike vectors lying within 
the  ``cone'' of opening angle $\epsilon<<1$.
These vectors define the time axis of the two
rest frames\footnote{If relative to these frames 
the inequality
$\tau_{R}(p)<<\tau_{C}(p)$
holds for all $p$ within the fluids interior, then 
these class of 
rest frames is refereed
as the admissible class of rest frames, a term adapted from the terminology
employed in transient thermodynamics. Local Lorentz transformations between such frames are
approximately described by 
(\ref{FC}).} and 
by complementing them by two triads
of spacelike unit vectors $e_{i}, i \in (1,2,3)$ and
$\hat e_{i}, i \in (1,2,3)$, then 
$(u, e_{i})$ respectively 
$(\hat u, \hat e_{i})$ 
constitute an orthonormal bases at the event under consideration.
Accordingly, we have
 
\begin{equation}
	\hat u=\frac {u}{(1-\frac {v^{2}}{c^{2}})^{\frac{1}{2}}}+ \frac {v^{i}e_{i}}{c}\frac {1}{(1-\frac {v^{2}}{c^{2}})^{\frac{1}{2}}},\quad
	v^{2}=v^{i}v_{i}.	
	\label{NVR}
\end{equation}  
  where $v^{i}$ are the components of the three velocity of the frame $u$ relative to the $\hat u$ one.
Following Israel \cite{Isr1},
we write
this transformation law in the equivalent form
 \begin{equation}
	\hat u^\mu = (1+\hat \epsilon ^2)^{1/2}u^\mu +\hat \epsilon^\mu, \qquad \hat \epsilon ^2 = \hat\epsilon ^a \hat \epsilon_a, \qquad \hat \epsilon ^a u_a = 0.
\label{VR}
\end{equation}}
and assume  $\hat \epsilon^{\mu} \leq \epsilon^{\mu}=O_1 $
 so that (\ref{VR})
is approximated by 
 \begin{equation}
\hat u^{\mu}=u^{\mu}+\hat{\epsilon}^{\mu}+O(\hat{\epsilon})^{2},\quad  \hat{\epsilon}^{\mu} \leq O_1, 
\label{FC}
\end{equation}
showing that
 $\hat{\epsilon}$  is a measure of the relative three velocity
of 
 $u^{\mu}$
relative to  $\hat u^{\mu}$ frame.\\  
 
%Since both $u^{\mu}$ and $\hat u^{\mu}$  lie within the
%cone of the opening angle $\epsilon$
%and furthermore it is assumed
%that the inequality $\tau_{C}(p)<<\tau_{M}(p)$ holds,
%then we may consider
Let now the 
 local ''thermodynamical equilibrium state'' associated
with  $(u^{\mu}, s(n,\rho))$  constructed 
according to (\ref{23}, \ref{33}).
If 
$Z(u)$ stands for any of the thermodynamical variable
in
(\ref{23})
(as measured by the $u$ observer)
and we consider the 
frame change described in (\ref{FC}),
of relevance for the following analysis 
is the 
 variation
 $\delta Z:=Z(\hat u)-Z(u)$ that suffers the variable $Z$ under such frame change.
Variations of these type worked out  by Israel in \cite{Isr1} and also reworked in \cite{Ftes}
and below we present a summary of such variations:

\begin{align}
	\delta \rho & \equiv \rho(\hat{u}) - \rho(u) = \hat{\epsilon}O_1 \label{lema1_02} \\[10pt]
	\delta h^\alpha & \equiv h^\alpha(\hat{u}) - h^\alpha(u) = -(\rho +P)\hat{\epsilon}^\alpha \label{lema1_03} \\[10pt]
	\delta n & \equiv n(\hat{u}) - n(u) = \hat{\epsilon}O_1, \label{lema1_04} \\[10pt]
	\delta P & \equiv P(\hat{u}) - P(u) = \hat{\epsilon}O_1, \label{lema1_05} \\[10pt]
	\delta \tau^{\alpha \beta} & \equiv \tau^{\alpha \beta}(\hat{u}) - \tau^{\alpha \beta}(u) = \hat{\epsilon}O_1 	\label{lema1_06}. \\[10pt]  
	\delta  n^\alpha & \equiv {n}^\alpha(\hat{u}) - {n}^\alpha(u) =-n \hat{\epsilon}^\alpha,
	\label{lema3_09}	
\end{align}
\begin{align}
        	\delta s & \equiv s(\hat{u}) - s(u) = \hat{\epsilon}O_1, \label{lema3_06} \\[10pt]
		\delta T & \equiv T(\hat{u}) - T(u) = \hat{\epsilon}O_1, \label{lema3_07} \\[10pt]
			\delta \Theta & \equiv \Theta(\hat{u}) - \Theta(u) = \hat{\epsilon}O_1. \label{lema3_08} 
	\end{align}
%\end{theorem}
These transformation shows
 that the variables like $(n,  \rho, P, etc)$
 behave as frame independent  quantities to 
 an accuracy $\hat \epsilon O_{1}\leq O_{2}$ while others variables like
$h^{\alpha}, n^{\mu}$ are 
independent only  to $\hat \epsilon\leq O_{1}$ accuracy\footnote{For the particular case where $ (u^\mu, \hat{u}^\mu) $ are identified as $ (u^\mu_E, u^\mu_N)$, the transformation law in (\ref{VR}) take the form:
  \begin{equation}
  u^{\mu}_{N} \to {u}^{\mu}_{E}=u^{\mu}_{N}+{\epsilon} ^{\mu}+O(\epsilon)^{2}
\label{ECTLL}
 \end{equation}
then one replaces in  
 the right hand sides of
 (\ref{lema1_02}-\ref{lema3_08})
 $\hat \epsilon$ by 
 $\epsilon=O_{1}$ 
 and the resulting  formulas 
 describes frame change from the particle to the energy frame.} .\\
For later use, we mention
that the following $4$-vector

\begin{equation}\label{lema2_01}
	q^\alpha(u) = h^\alpha(u) - \frac{\rho(u) + P(u)}{n(u)}n^\alpha(u),\\[10pt]
\end{equation}
is frame independent i.e. its variation obeys:
\begin{equation}\label{lema2_02}
\delta q^\alpha \equiv q^\alpha(\hat{u}) - q^\alpha(u) = \hat \epsilon^ {\alpha}O_1. \\[10pt]
\end{equation}
and is refereed as the 
the invariant heat flux vector.\\

The formulas in
(\ref{lema1_02}-\ref{lema3_08})
express the 
transformation laws 
of the various thermodynamical variables 
under a frame change described in 
(\ref{FC})
and there will be used frequently in the following analysis.
%They permit us  to develop 
% the thermodynamics
%of small deviations from the state of
 %''local ''thermodynamical equilibrium'' 
%specified for instance  by $(u^{\mu}, s(\rho, n))$. 
Under such  frame 
 change, they 
 show that  most of the thermodynamical variables 
 remain practically frame independent as long as second order and higher order deviations from the
 state of 
 ``local thermodynamical equilibrium'' 
specified by  $(u^{\mu}, s(\rho, n))$,  are
neglected. Notice that the assumption
 $\epsilon<<1$ in (\ref{PSA}) 
implies that
the inequality 
$\tau_{C}(p)<<\tau_{M}(p)$,
becomes observer independent 
in the sense that 
as long as 
 $(u^{\mu}, \hat u^{\mu}) $
in 
(\ref{FC}) are chosen to lie within the ``cone''
 of the opening 
  pseudo-angle 
$\epsilon = O_{1}<<1$ 
then time dilatation and length contraction are 
 effects considered as been inessential.
 Moreover, the ''invariance'' of the inequality 
$\tau_{C}(p)<<\tau_{M}(p)$,
under the  frame 
change in 
(\ref{FC}) 
in combination to  (\ref{lema1_02}-\ref{lema3_08}),
permit us to simply refer to a state 
 of ``local thermodynamical equilibrium'' 
without any further reference to which particular $(u^{\mu}, s(\rho, n))$
this fictitious state'
is associated with.\\
 
 The so far analysis 
used only the property that fluid states
 are described (partially) by
  the fields $(T^{\mu\nu}, J^{\mu})$
  and all of the above conclusions are independent of any underlying fluid theory.
  This observation 
  allows us  to group together states 
  characterized by common properties.   
  To do so, 
  let us begin with a state
   described by 
  the fields $(T^{\mu\nu}, J^{\mu})$ 
subject to the restriction
  that the pseudo-angle 
   $\epsilon$ in (\ref{PSA}) satisfies
 $\epsilon<< 1$ everywhere within the fluid region   
  and let
   a four velocity $u^{\mu}$ 
lying within the cone of opening angle $\epsilon<<1$
chosen so that
$\tau_{C}(p)<<\tau_{M}(p)$
holds.     
   Any such $u^{\mu}$ combined with  
$(T^{\mu\nu}, J^{\mu})$ defines a particular fluid state
which is described by the fields defined in the expansions 
(\ref{OD}, \ref{OD_11}).
We refer to the collection of such states, as states satisfying the relativistic (LTE) postulate (or often
as states near equilibrium).\\   

One notices that for any two states within this class, specified 
by  $u^{\mu}$, respectively 
 $\hat u^{\mu}$ 
 it holds $\hat u^{\mu}- u^{\mu}\leq \epsilon ^{\mu}$
 and thus  under a frame change
 described by
(\ref{FC}), 
 the fields 
 in (\ref{23}, \ref{H22})
transform according to
 (\ref{lema1_02}-\ref{lema3_08}).
Therefore even though technically one deals with the two distinct fluid states,
as long as one is interested only in the physics of first order deviations from the state of 
''local thermodynamical equilibrium''
 one really is dealing with the same state.
Moreover in the next sections, we show that 
for a large class of fluid theories, states 
 compatible with this relativistic (LTE) postulate,
   the phenomenological equations
 that describe the dynamics of first order deviations from the 
 ''local thermodynamical equilibrium'' state
are equivalent in the sense that from one solution  one generates solutions of 
the other equation (or equivalently from one state in (LTE) one generates all the other states within this class).\\

As a preparatory step to analyze further properties of 
states satisfying the relativistic (LTE) postulate,
let us consider one of them described  by
$(T^{\mu\nu}, J^{\mu})$ and let us introduce also a physical entropy current $S^{\mu}$
associated to this state.
If and in accordance of the above discussion, this state is
specified by 
$(u^{\mu}, s(\rho, n))$,
we define the fields, 
\begin{equation}
\delta S^{\mu}
=S^{\mu}-S_{0},\quad
\delta T^{\mu\nu}= T^{\mu\nu}- T^{\mu\nu}_{0},\quad
\delta J^{\mu}=
J^{\mu} - J^{\mu}_{0}
\label{44}
\end{equation}
which describe  the deviations of the physical state $(S^{\mu},T^{\mu\nu}, J^{\mu})$
away from the state of ''local thermodynamical equilibrium'' state defined
(by
 $(u^{\mu},s(\rho, n))$)
or by
$(S^{\mu}_{0}, T^{\mu\nu}_{0}, J^{\mu}_{0})$ 
as defined in 
according to (\ref{33}).\\
Since by construction,
$( T^{\mu\nu}_{0}, J^{\mu}_{0})$
satisfy the fitting conditions
 \begin{equation}
(J^{\mu}-J_{(0)}{}^{\mu})u_{\mu}=(T^{\mu\nu}-T_{(0)}{}^{\mu\nu})u_{\mu}u_{\nu}=0,	
\label{FC1}
\end{equation}
the perturbations $\delta J^{\mu}$ and 
$\delta T^{\mu\nu}$ are described by 
\begin{equation}
\delta J^{\mu} = n(u)^{\mu}
\end{equation}
\begin{equation}
\begin{split}
\delta T^{\mu\nu} & = h(u)^{\mu}u^{\nu}+h(u)^{\nu}u^{\mu}+\tau(u)^{\mu\nu}\\
&
 = \pi(u)\Delta(u) ^{\mu\nu}+\pi(u)^{\mu\nu}+
 h(u)^{\mu}u^{\nu}+h(u)^{\nu}u^{\mu},
\end{split}
\end{equation}
and these forms are actually independent of the dynamics of the underlying fluid theory.
However that is not any longer the case for the entropy perturbation
$\delta S^{\mu}
=S^{\mu}-S^{\mu}_{0}$, 
the structure of $S^{\mu}$ plays an important role  in specifying  he underlying theory.\\

Before we leave this section, it is worth 
stressing a point. For states satisfying the relativistic (LTE) postulate,
specified by
$(u^{\mu}, s(\rho, n))$
even though 
the equilibrium equation of state
$s(\rho, n)$ 
combined with
the velocity field $u^{\mu}$
defines an entropy like  current
$S^{\mu}_{0}=s(\rho, n)u^{\mu}$ as defined in (\ref{33})
this current is formal and has nothing to do with the 
 physical entropy 
$S^{\mu}$ of the underlying state.
Of relevance in the 
analysis of relativistic fluid states is the dependance 
of the physical entropy current $S^{\mu}$ 
upon other fluid variables
and below, we shall have the opportunity to see 
how this dependance
leads to alternative theories of relativistic dissipation.\\

 \section{On states satisfying the relativistic (LTE) postulate and  first order theories}

In this section, we study states compatible with the relativistic (LTE) postulate
 within the context of the Hiscock-Lindblom class of first order theories\footnote{The term 
  ''first order theories''  coined by 
 Hiscock and Lindblom in \cite{His2} \cite{His3}) and describes theories where the entropy current $S^{\mu}$ 
 receives only first order contributions from a suitably defined state of ''local thermodynamical equilibrium''.
 Initially in this section we shall be concerned with the Hiscock-Lindblom class, but at the end of the section we shall introduce the (BDNK) class of first order theories that differs from the Hiscock-Lindblom class.}.
We recall first that this class has been introduced  in \cite{His2}  \cite{His3}  and 
%This is a large class of theories which includes the Landau-Lifshitz theory and Eckart theory as particular cases.\\
 states within this class,
 are described by the  fields
$(T^{\mu\nu}, J^{\mu}, S^{\mu})$ satisfying
(\ref{BE},\ref{EL}). Even though states within the theory do not
 make no reference to any fluid four velocity, nevertheless, in all treatments 
of this theory, a velocity
field $u^{\mu}$ creeps in (see for instance
the analysis  
of Lindblom and Hiscock in \cite{His2} \cite{His3}). Once a choice of 
a four velocity\footnote{Just to avoid confusion, we stress that the choice of this velocity field  $u^{\mu}$ is arbitrary,
 (except that it is restricted so that relative to the family of the rest frames that it defines, the inequality
$\tau_{C}(p)<<\tau_{M}(p)$
holds)
and this
$u^{\mu}$
bears no relation to the four velocity field
entering in the specification of states compatible with the (LTE) postulate. For instance, we have not yet introduced 
the pseudo angle $\epsilon$ defined in (\ref{PSA}) restricted to obey $\epsilon<< 1$.} $u^{\mu}$ has been made, 
it is implicitly assumed that
relative to the family of rest frames defined by 
this 
$u^{\mu}$, the inequality
 $\tau_{C}(p)<<\tau_{M}(p)$
holds and thus local thermodynamical equilibrium prevails.
 In turn this allows one to postulate the existence of an ''equilibrium equation of state'' $s$
which combined with 
$(T^{\mu\nu}, J^{\mu}, S^{\mu})$ 
defines 
a ''local thermodynamical equilibrium'' state attached to 
$(u^{\mu},s(\rho, n))$
in the manner
discussed in the previous section.
%The Landau-Lifshitz  theory \cite{LaL} is identified as the first order theory where $u^{\mu}$ is
%chosen to be $u^{\mu}_{E}$,while for the Eckart theory \cite{Eck}
%is identified by choosing $u^{\mu}$ to be $u^{\mu}_{N}$.
\\
A key ingredient that characterizes the Hiscock-Lindblom class
is the structure of  the  physical entropy current 
$S^{\mu}$ which is postulated to have the form (see discussion in \cite{His2})
\begin{equation}
S^{\mu}=su^{\mu}+\hat \beta h^{\mu}(u)-\hat \Theta n^{\mu}(u)=S^{\mu}_{0}+\hat \beta h^{\mu}(u)-\hat \Theta n^{\mu}(u),
\label{EFO}
\end{equation}
where $(\hat \beta, 
\hat \Theta)$ are undetermined functions.
The choice
$(u^{\mu}=u^{\mu}_{E}, h^{\mu}(u)=0)$ 
generates the
Landau-
Lifshitz theory, while 
$(u^{\mu}=u^{\mu}_{N}, n^{\mu}=0)$ generates the 
Eckart theory.\\
With reference to the velocity field
$u^{\mu}$ in
(\ref{EFO}),
 the 
expansions 
of $(T^{\mu\nu}, J^{\mu})$
 in
(\ref{OD},\ref{OD_11}),
introduce the fields 
$(\rho(u), n(u), P(u), h^{\mu}(u), \tau^{\mu\nu}(u), n^{\mu}(u))$
which are required to obey:

\begin{equation}
\begin{split}
\nabla_{\mu}T^{\mu\nu}=\nabla_{\mu}[& \rho(u)u^{\mu}u^{\nu}+P(u)\Delta(u) ^{\mu\nu}+h(u)^{\mu}u^{\nu} \\
& + h(u)^{\nu}u^{\mu}+\tau(u)^{\mu\nu}]=0,
\label{FOD}
\end{split}
\end{equation}

\begin{equation}
	\nabla_{\mu} J^{\mu}=\nabla_{\mu}[n(u)u^{\mu}+n(u)^{\mu}]=0.
\label{FOD_11}	
\end{equation}
Moreover,  by imposing the second law 
 $\nabla_{\mu}S^{\mu}\geq 0$  and recalling the decomposition
$\tau(u)^{\mu\nu}=\pi(u)\Delta(u) ^{\mu\nu}+\pi(u)^{\mu\nu}$, $\pi(u)^{\mu}_{\mu}=0$
in (\ref{SD}),
 a calculation shows that this law can be fulfilled whenever the following equations hold
(for a derivations see \cite{His2}):

\begin{equation}
h^{\mu}(u)=-kT(u)\Delta^{\mu\nu}(u) [\frac {1}{T(u)} \nabla_{\nu}T(u)+u^{\alpha} \nabla_{\alpha} u_{\nu}],
\label{1}	
\end{equation}

\begin{equation}
n^{\mu}(u)=-\sigma T^{2}(u) \Delta(u)^{\mu\nu}\nabla_{\nu} \Theta(u),
\label{2}	
\end{equation}

\begin{equation}
\pi^{\mu\nu}(u)=-2 \eta <\nabla^{\mu}
u^{\nu}>,
\label{3}	
\end{equation}

\begin{equation}
\quad \pi(u)=-\zeta\nabla_{\mu}u^{\mu},
\label{4}	
\end{equation}
with the coefficients  $(\hat \beta, 
\hat \Theta)$  
in (\ref{EFO}) given by:

\begin{equation}
\hat \beta(u)=T^{-1}(u):=\beta(u),\hspace{0.15cm} \hat \Theta(u)=\frac {\rho(u)+P(u)}{n(u)T(u)}-s(u):=\Theta(u),
\label{5}	
\end{equation}
and the  angular bracket in (\ref{3}), and here after, signifies symmetric, purely spatial, trace free part of the enclosed
tensor. From these equations
one gets

\begin{equation}
T\nabla_{\mu}S^{\mu}=\frac {\pi^{2}}{\zeta}+\frac {h^{\mu}h_{\mu}}{kT}+\frac {n^{\mu}n_{\mu}}{\sigma T}+
\frac {\pi^{\mu\nu}\pi_{\mu\nu}}{ 2 \eta}\geq0,
\label{66}	
\end{equation}

implying that the entropy production 
 is manifestly non negative provided the four coefficients $(\zeta, \eta, k, \sigma)$
are chosen  to be positive and these 
coefficients are identified as 
the bulk viscosity $\zeta$, the thermal conductivity $k$, a particle diffusion 
constant $\sigma$
and the shear viscosity $\eta$.\\

The system
(\ref{FOD}-\ref{5})
constitutes a closed system of equations
whose solutions describe arbitrary fluid states within 
the Hiscock-Lindblom class of first order theories.
From the mathematical view point, this system  is a mixed
parabolic-hyperbolic-elliptic system and as we shall discuss further ahead,
its solutions are characterized by a number of undesirable properties. Nevertheless, viewed as a closed system, 
and given 
suitable initial data,
determines
the unknowns variables
and solutions 
possessing distinct
velocity fields
are considered to be distinct solutions.\\

However,
matters differ
 when emphasis is restricted to states satisfying the relativistic (LTE) postulate. 
Primary for such states,
the fields $T^{\mu\nu}$ and $J^{\mu}$
are restricted so that
 the pseudo-angle 
$\epsilon$ 
between 
$u_{E}$ and  $u_{N}$ 
(see (\ref{PSA})) 
satisfies
everywhere
within the region occupied by the fluid
the condition
$\epsilon<<1$ and moreover 
exists a (highly non unique) velocity 
field
$u^{\mu}$ that 
lies within the cone of the opening angle 
$\epsilon<<1$ that generates a 
state of the ''local thermodynamical equilibrium''
state specified by $(u^{\mu}, s(\rho, n))$
in the manner discussed  in the previous section.
Clearly there exist infinitely many states satisfying the (LTE) postulate, with each one of them is specified by
the same
pair $(T^{\mu\nu}, J^{\mu})$
but  by a different velocity field.\\
We show below, that all 
such states 
are in effect equivalent to each other
as long as quadratic 
and higher order deviations from the state of the ''local thermodynamical equilibrium''
specified by $(u^{\mu}, s(\rho, n))$
are omitted.\\
%  and the
%gradients of $h^{\mu}$, $\pi$, and $\pi^{\mu\nu}$
%are to be depreciated in comparison to the gradients of
%the velocity field.
To establish this equivalence, let us begin with two velocity fields
$(u^{\mu}, \hat u^{\mu})$ that specify two states satisfying the relativistic (LTE) postulate. 
Using $u^{\mu}$,
let  
\begin{equation}
(n( u),  \rho(u),  P(u),  h(u)^{\mu},  n(u)^{\mu}, \tau(u)^{\mu\nu})
\label{6}	
\end{equation}
be the fields obtained by expanding
$(T^{\mu\nu}, J^{\mu})$ 
according to (\ref{OD},\ref{OD_11}) 
and let us assume that these fields
satisfy the exact equations	
(\ref{FOD}-\ref{5})
for a set on non negative coefficients $(\zeta, \eta, k, \sigma)$.
Using this solution 
as a reference, we shall generate all other states compatible with the relativistic (LTE) postulate.
For this, let the 
 frame change
\begin{equation}
u^{\mu}\to \hat u^{\mu}=u^{\mu}+\hat \epsilon^{\mu},\quad \hat\epsilon \leq \epsilon,
\label{FRCH}	
\end{equation}
then 
(\ref{lema1_02}-\ref{lema3_08})
combined with the fields 
in (\ref{6}), generate
$$(\hat u^{\mu}, \rho(\hat u), n(\hat u), P(\hat u), h^{\mu}(\hat u), \tau^{\mu\nu}(\hat u), n^{\mu}(\hat u)).$$
The claim is, that this new state
 satisfy 
(\ref{FOD}-\ref{4}), as long as deviations
from a the state of ''local thermodynamical equilibrium''
specified by $(u^{\mu}, s(\rho, n))$
are omitted.\\
 
In order to
show this, we first consider the special case 
where 
$\hat u^{\mu}$
is chosen to be 
the four velocity $u^{\mu}_{E} $ of the energy frame 
so that 
(\ref{FRCH}) takes the form	
\begin{equation}
u^{\mu} \to u^\mu_{E} =u^{\mu}+\hat \epsilon ^{\mu}, \quad
	\hat \epsilon^{\mu} \leq \epsilon^{\mu}. 
	\label{VR122}
\end{equation}
Using
 (\ref{lema1_02}-\ref{lema3_08}),
 let 
\begin{equation}
  u^\mu_{E}, n(u_{E}),  \rho(u_{E}),  P(u_{E}),  h( u_{E})^{\mu}=0, n(u_{E})^{\mu}, \tau(u_{E})^{\mu\nu}
\label{FOVA}	
\end{equation}
defined according to
$$ n(u_{E})= n( u)+\hat \epsilon O_{1},\quad \rho(u_{E})= \rho( u)+\hat \epsilon O_{1},\quad etc,$$
and in this generation process, it is worth noticing that  
the condition
$h^{\mu}(u_{E})=0$ demands that  $\hat \epsilon^{\mu}$ to satisfy
\begin{equation}
	h^\mu({u}) =(\rho+P)\hat \epsilon^{\mu},
		\label{VR12}
\end{equation}
while
$n^{\mu}(u_{E})$ has the the value

\begin{equation}
n^{\mu}(u_{E})=n^{\mu}(u)-n \hat \epsilon^{\mu}=n^{\mu}(u)-\frac {n(u) h^{\mu}(u)}{(\rho+P)}.
\label{VR13}
\end{equation}

The state in (\ref{FOVA}), approximately  satisfy 	
(\ref{FOD}-\ref{5}). Indeed, starting from eqs.(\ref{FOD})
and replacing $u^{\mu}$
by 
$u^{\mu}_{E}$ and 
$(n( u),  \rho(u),)$ etc
by the fields in 
(\ref{FOVA}), we find
 
\begin{equation}\label{EQUI1} \begin{split}
 0 = &\nabla_{\mu}T^{\mu\nu}(u) \\
  = & \nabla_{\mu}[\rho(u)u^{\mu} u^{\nu}+P(u)\Delta(u)^{\mu\nu}+h(u)^{\mu}u^{\nu}+h(u)^{\nu}u^{\mu}+\tau(u)^{\mu\nu}] \\
  = & \nabla_{\mu}[\rho(u_{E})u^{\mu}_{E} u^{\nu}_{E}+P(u_{E})\Delta(u_{E}) ^{\mu\nu}+ \tau(u_{E})^{\mu\nu}
+O(\epsilon O_{1})],
 \end{split}
\end{equation} 
 
 where we arrived at the last equality using $u^{\mu}=u^{\mu}_{E} -\hat \epsilon^{\mu}+O(\hat \epsilon)^{n}, n\geq 2$ and
 $h^{\mu}(u)=(\rho + P)\hat \epsilon^{\mu}+
 O(\hat \epsilon)^{n}, n\geq 2$.
 %$ n^{\mu}(u)=n^{\mu}(u_{E})+n \hat \epsilon^{\mu}$. 
 Thus
the state in 
(\ref{FOVA})	
satisfy
the conservation equation $\nabla_{\mu}T^{\mu\nu}=0$, provided terms of 
order $(\epsilon O_{1})$ and their gradients are thrown away.
Via a  similar reasoning, but now starting from 
eq. (\ref{FOD_11}), one finds

\begin{equation}
\begin{split}
	0 & = \nabla_{\mu} J^{\mu}=\nabla_{\mu}[n(u)u^{\mu}+n(u)^{\mu}] \\
	& =
\nabla_{\mu}[n({u_{E}})u^{\mu}_{E}+n^{\mu}(u_{E})+O(\epsilon O_{1})],	
\label{FOD_112}	
\end{split}
\end{equation}
where we used
$n^{\mu}(u_{E})=n^{\mu}(u)-n \epsilon^{\mu}
+O(\hat \epsilon)^{n}, n\geq 2$ and thus  to an $O_{1}$ accuracy
the state in 
(\ref{FOVA})	
indeed satisfy the conservation law in (\ref{FOD_11}).\\
 
 We now examine 
 % turn attention to equations (\ref{1},\ref{4}) and check %
 whether the fields $( n(u_{E})^{\mu}, \tau(u_{E})^{\mu\nu})$
satisfy to linear order
equations (\ref{2}-\ref{4}).
 The easiest way to prove this assertion is
to start 
from the entropy current $S^{\mu}$ in (\ref{EFO}) and eliminate
$u^{\mu}, h^{\mu}(u)$ and $n^{\mu}(u)$ 
in favor of the fields measured relative to the energy frame.
Using
$u^{\mu}=u^{\mu}_{E} -\hat \epsilon^{\mu}+O(\hat \epsilon)^{n}, 
 h^{\mu}(u)=(\rho + P)\hat \epsilon^{\mu}+
 O(\hat \epsilon)^{n}, n\geq 2$,
it follows  that 
$S^{\mu}$
takes the form
\begin{equation}\label{EFO1}
\begin{split}
S^{\mu} & = s(u^{\mu}_{E}-\hat \epsilon^{\mu})+\beta (\rho+P)\hat \epsilon^{\mu}
-\Theta (n^{\mu}(u_{E})+n\hat \epsilon^{\mu}) \\
 & = su^{\mu}_{E}-\Theta(u)n^{\mu}_{E}-[s-\beta(\rho+P) +\Theta n]\hat 
\epsilon^{\mu}+O(\epsilon^{2}) \\
 & = s(u_{E})u^{\mu}_{E}-\Theta(u_{E})n^{\mu}_{E}+O(\epsilon^{2}) 
\end{split}
\end{equation}
where
$s(u_{E}):=s(\rho(u_{E}), n(u_{E}))$
and we used the fundamental relation $s=(\rho+P)T^{-1}-\Theta n$.
By imposing the second law on this form 
of $S^{\mu}$ but now written relative to the energy frame,
one arrives at
\begin{equation}
n^{\mu}(u_{E})=-\sigma T^{2} \Delta(u_{E})^{\mu\nu}\nabla_{\nu} \Theta(u_{E})+O(\hat \epsilon)^{2},
\label{2E}	
\end{equation}

\begin{equation}
\pi^{\mu\nu}(u_{E})=-2 \eta <\nabla ^ {\mu}u^{\nu}_{E}+O(\hat \epsilon)^{2}>,
\label{3E}	
\end{equation}

\begin{equation}
\quad \pi(u_{E})=\zeta\nabla_{\mu}u^{\mu}_{E}+O(\hat \epsilon)^{2}.
\label{4E}	
\end{equation}

Thus the
state in 
(\ref{FOVA})
satisfy the equations of
the Landau-Lifshitz theory provided that 
non linear terms in the deviations from the 
state of ''local thermodynamical equilibrium'' 
and gradients of  $h^{\mu}$, $\pi$, and $\pi^{\mu\nu}$
have been depreciated\footnote{
Notice that
equations (\ref{3E}-\ref{4E})
 could be derived directly from 
(\ref{3},\ref{4}) by replacing 
$u^{\mu}$ by
$u^{\mu}=\hat u^{\mu}-\epsilon^{\mu}$
and treating  $(\nabla_{\mu}\epsilon ^{\mu}, \nabla_{\mu} \epsilon_{\nu})$
 as $\epsilon O_{1}$ and thus depreciated them  in comparison to the gradients of the velocity
 field (this estimate will be employed further ahead)}.

It is worth noticing that if in this analysis,
 we replace $u^{\mu}$  by $u^{\mu}_{N}$
and  
set
$n^{\mu}(u_{N}):=0$ 
in 
(\ref{6}) and  $\sigma :=0$ in 
(\ref{2}), so that 
	
$$u^{\mu}_{N}, n(u_{N}),  \rho(u_{N}),  P(u_{N}),  h^{\mu}( u_{N}), \tau^{\mu\nu}(u_{N}),$$  
satisfy
the exact equations
(\ref{FOD}-\ref{5})
for the Eckart theory,
then
the state
$u^{\mu}_{E}, n(u_{E}),  \rho(u_{E}),  P(u_{E}),  n(u_{E})^{\mu}, \tau(u_{E})^{\mu\nu}$
satisfy approximately the equations for 
the Landau-Lifshitz theory. This process can be also reversed i.e. starting from a state
satisfying the
the Landau-Lifshitz theory, one can generate a state that to 
 $O_{1}$ accuracy satisfy
the Eckart  theory.\\
%Thus at the linear order in the deviations, states satisfying the relativistic (LTE) postulate,
%the Landau-Lifshitz and Eckart theories are equivalent theories.
Actually a more general 
statement holds. Within the 
context of the Hiscock-Lindblom class of
first order theories,
states compatible with the relativistic (LTE) postulate
are to linear order equivalent states. To verify this claim, we return to the original  fields  
 $(u^{\mu}, \hat u^{\mu})$ and consider the  
 frame change in (\ref{VR122}). We assume
that $u^{\mu}$,
and the  fields in (\ref{6}) satisfy (\ref{OD},\ref{OD_11}) 
and generate 
$n(\hat u),  \rho(\hat u),  P(\hat u),  h(\hat u)^{\mu},  n(\hat u)^{\mu}, \tau(\hat u)^{\mu\nu}$
according to
$$ n(\hat u)= n( u)+\hat \epsilon O_{1},\quad \rho(\hat u)= \rho( u)+\hat \epsilon O_{1},\quad etc.$$
By similar arguments
that lead us to 
(\ref{EQUI1},\ref{FOD_112}), the conservation eqs yield

\begin{equation}\label{LMR011}
\begin{split}
0 & = \nabla_{\mu}T^{\mu\nu}\\
& = \nabla_{\mu}[(\rho+P)\hat u^{\mu}\hat u^{\nu} + P g^{\mu\nu}+h(\hat u)^{\mu}\hat u^{\nu} + h(\hat u)^{\nu}\hat u^{\mu} \\
& + \tau(\hat u)^{\mu\nu} +\hat \epsilon O_{1}] - \nabla_{\mu}[h^{\mu}(\hat u)\epsilon^{\nu} + h^{\nu}(\hat u)\epsilon^{\mu}] 
\end{split}
\end{equation} 

\begin{equation}\label{LMR022}
\begin{split}	
	0  & = \nabla_{\mu} J^{\mu}=\nabla_{\mu}[n(u)u^{\mu}+n(u)^{\mu}]\\
	& =
\nabla_{\mu}[n(u)\hat u^{\mu}+n(\hat u)^{\mu}+\hat \epsilon^{\mu}O_{1}],
\end{split}
\end{equation}
and  under the assumptions that gradients of $h^{\mu}$ and of $\epsilon^{\mu}$
are neglected in comparison to the velocity gradients, it follows that 
the terms linear in $\epsilon^{\mu}$ in the right hand side of 
(\ref{LMR011}, \ref{LMR022})
can be neglected and thus 
the conservation laws (\ref{FOD}, \ref{FOD_11})
to an $O_{1}$ accuracy
holds true
for the state
 $(\hat u^{\mu}, n(\hat u),  \rho(\hat u),  P(\hat u),  h(\hat u)^{\mu},  n(\hat u)^{\mu}, \tau(\hat u)^{\mu\nu})$.\\
We now indicate that 
 equations 
 (\ref{1}-\ref{4}) remain valid and for this
 we observe that the entropy flux 
 $S^{\mu}$
 in (\ref{EFO})
 that under 
 the frame change described in (\ref{VR122}) implies:	
 
 \begin{equation}
 \begin{split}
S^{\mu} & =
su^{\mu}+\beta(u) h^{\mu}(u)-\Theta (u) n^{\mu}(u) \\
& =
s\hat u^{\mu}+ \beta(\hat u) h^{\mu}(\hat u)-\hat \Theta (u) n^{\mu}(\hat u) +O(\epsilon^{2})
\end{split}
\label{EFO2}
\end{equation} 
and this  $O_{1}$-invariance property of 
 $S^{\mu}$
 implies  
that the fields
$h(\hat u)^{\mu},  n(\hat u)^{\mu}, \tau(\hat u)^{\mu\nu}$
satisfy the analogues of 
(\ref{1}-\ref{4}) to an $O_{1}$ accuracy i.e to linear order in the deviations from
the state of 
''local thermodynamical equilibrium'' specified by $(u^{\mu}, s(\rho, n))$.

Before we continue, we offer a few comments regarding the 
constitutive relations 
that states satisfying the relativistic (LTE) postulate obey.  
 We recall first the derivation 
of the exact 
relations (\ref{1}-\ref{4}). For this derivation, one begins by imposing the second law on the the entropy current $S^{\mu}$ in (\ref{EFO})
and using the exact eqs $ \nabla_\mu T^{\mu \nu} = \nabla_\mu J^\mu  = 0 $ one 
arrives at
the exact relations (\ref{1}-\ref{4}) with the positive 
 coefficients $(\zeta, \eta, k, \sigma)$ putting in by hand.
 However for states satisfying the (LTE) postulate, the constitutive relations obeyed by $(h(u), n^{\mu}(u), \pi^{\mu\nu}(u), \pi(u))$
 hold  approximately, since
  in their derivation we used the approximate expressions for the entropy currents $S^{\mu}$ shown 
 in the last lines of (\ref{EFO1}) and (\ref{EFO2}). It is worth however, to probe this approximation
 and in particularly to probe the transition from the constitutive relations of
 the  Eckart theory  to the corresponding relations for the Landau-Lifshtz theory. To do so,
 we
 start from the 
easily verifiable identity (see for instance eq.$(34)$ in \cite{His1})
 
 \begin{equation}
nT\nabla_\alpha \Theta = \nabla_\alpha P - \frac{\rho + P}{T}\nabla_\alpha T
\label{A0}
\end{equation}  
and let for generality  purpose assume that the fluid is described by a $T^{\mu\nu}$ shown  in 
(\ref{OD}).
%energu mommentim tensor as in
%for a $ T^{\mu \nu} = (\rho + P) u^\mu u^\nu + P\Delta^{\mu \nu} + h^\mu u^\nu + h^\nu u^\mu + \tau^{\mu \nu} $
For this case the following identity holds

\begin{equation}
\Delta^{\mu \nu}\left[\frac{\nabla_\nu T}{T} + a_\nu\right] = - \frac{nT}{\rho + P}\Delta^{\mu \nu}\nabla_\nu \Theta + \frac{F^\mu}{\rho + P},\quad a^{\mu}=u^{\nu}\nabla_{\nu}u^{\mu}
\label{A1}
\end{equation}
where $F^{\mu}$ is defined from the conservation equation
$\Delta^{\mu}{}_{\rho}\nabla_{\nu}T^{\nu\rho}=0$ which can be written in the equivalent form
\begin{equation}
\Delta^{\mu \nu}\nabla_\nu P + (\rho + P)a^\mu = F^\mu 
 \label{A2}
\end{equation}
where the precise form of $F^{\mu}$ is easily derivable, but it is not really needed
since it contributes terms of
$O(\epsilon^{2})$ or terms that make negligible contribution like $\epsilon^{\mu}\nabla_{a}u^{a}$ etc.
 It is understood that
 in 
 (\ref{A0}-\ref{A2}), the terms 
 $T, \Theta, P, \Delta^{\mu\nu}$  
 stand for
 $T(u), \Theta(u), P(u), \Delta^{\mu\nu}(u)$ etc.\\
 
 Suppose now we start from  the Eckart theory
so that the following relations hols
$$h^\mu(u_{N})  = -kT(u_{N}) \Delta^{\mu \nu}(u_{N})\left[\frac{\nabla_\nu T(u_{N})}{T(u_{N})} + a_\nu\right],\quad
\pi^{\mu \nu}(u_{N}) = -2n \langle \nabla^\mu u^\nu\rangle,\quad
\pi(u_{N}) = - \zeta \nabla_\mu u_{N}^\mu$$
for some non vanishing 
$(k, \zeta,\eta)$. 
Let us now perform a frame change
described by 
$$u_{N}^{\mu} \to u^\mu_{E} =u_{N}^{\mu}+\epsilon ^{\mu}$$ 
and for 
this frame change, we
evaluate the identity
(\ref{A1}) at $u_{N}^{\mu}$
and combine it  with  
\begin{equation}
h^\mu(u_{N})  = -kT(u_{N}) \Delta^{\mu \nu}(u_{N})\left[\frac{\nabla_\nu T(u_{N})}{T(u_{N})} + a_\nu\right].
 \label{A4}
\end{equation}
Using 
$h^{\mu}(u_{N})=(\rho+P) \epsilon^{\mu}$ ,$\quad n^{\mu}(u_{E})=-n\epsilon^{\mu}$ writing
$\Delta^{\mu\nu}(u_{N})= 
\Delta^{\mu\nu} (u_{E})-u_{E}^{\mu} \epsilon^{\nu}-u_{E}^{\nu} \epsilon^{\mu}+O(\epsilon^{2})$, 
$T(u_{N})=T(u_{E})+O(\epsilon^{2})$ etc, one 
finds after some algebra that
(\ref{A4}) transforms into
\begin{equation}
n^\mu(u_{E})  = -\sigma T^2(u_{E}) \Delta^{\mu \nu} (u_{E})\nabla_\nu \Theta(u_{E}) +O(\epsilon)
\label{A5}
\end{equation}

where  $O(\epsilon)$ denote terms involving products of $\epsilon$ with other other terms
which in general considered as negligible in comparison to the leading 
term $\sigma T^2(u_{E}) \Delta^{\mu \nu(u_{E}}\nabla_\nu \Theta(u_{E})$ and 
in arriving at 
(\ref{A5})
we introduced the coefficient  $ \sigma$ via  
  $$ \sigma =k \frac{n^2}{(\rho + P)^2} $$.\\ 
 
 Similarly, staring from
 $$\pi^{\mu \nu}(u_{N}) = -2n \langle \nabla^\mu u^\nu\rangle,\quad
 \pi(u_{N}) = - \zeta \nabla_\mu u_{N}^\mu,$$ 
 then 
 under the frame change
 $u_{N}^{\mu} \to u^\mu_{E} =u_{N}^{\mu}+\epsilon ^{\mu}$,
it can be easily seen that they transform into 
 $$\pi^{\mu \nu}(u_{E}) = -2n \langle \nabla^\mu u_{E}^\nu\rangle, \quad
 \pi(u_{E}) = - \zeta \nabla_\mu u_{u_{E}}^\mu,$$
 where we have dropped terms linear in $\epsilon^{\mu}$ and its derivatives.\\
 One can also reverse this process i.e. one can start from the
 constitutive relation in the 
  Landau-Lifshitz theory to generate the 
 constitutive relations for the Eckart theory.  
 More generally one
 can consider frame changes described for instance 
by  (\ref{FRCH}) and starting from the constitutive relations	 
 relative to the $u^{\mu}$ frame work out the
 constitutive relations	 
relative to the $\hat u^{\mu}$. In general one finds 
  to lowest order
(\ref{1}-\ref{4}) hold but they are  modified by corrections terms
 of order  $ O(\epsilon^\mu) $, etc. and these correction terms are neglected 
 as long as we are in the regime where $\epsilon<<1$
(see also  a discussion at  this approximation, in the review article by Israel in \cite{Isr5} page $(179))$.\\

%and terms of order $ (\epsilon O_1) $, $ \nabla_\mu(h^\mu \epsilon_1) $ etc.
%have been  are ignored. Therefore if one starts from 
%(\ref{1}-\ref{4}) and performs a change frame like the one shown in (51), then one would finds
%that to lowest order
%(\ref{1}-\ref{4}) hold but they are  modified by corrections terms
 %of order  $ O(\epsilon^\mu) $, etc. and these correction terms are neglected 
 %as long as we are in the regime where $\epsilon<<1$
%(see for instance the discussion at this point in the review article by Israel (\cite{Isr5} page $(179)$.
%In that regard it is worth to probe a bit the nature of the constitutive relations between
%the Eckart and Landau-Lifshtz theory. To make our point, let us start from the 
%easily verifiable identity (see for instance eq.$(34)$ in \cite{His1})

In summary, therefore within
 the Hiscock-Lindblom class of first order theories, states 
 that satisfy the relativistic 
 (LTE) postulate 
 are equivalent states as long as one neglects terms of $\epsilon^{2}$ and higher
order in the deviations from the local equilibrium field and 
gradients of $h^{\mu}$, $\pi$, and $\pi^{\mu\nu}$
are small compared to thermal and velocity gradients
and this conclusion holds also for the Landau-Lifshitz and Eckart 
theories.\\

In the so far analysis, we used the Hiscock-Lindblom class of first order theories as a test bed, to get insights on the properties of
states satisfying the relativistic (LTE) postulate. However, ought to be mentioned
that this class of theories, including the Eckart and 
Landau-Lifshitz theories,
are pathological. 
According to the results in \cite{His2, His3},
they do not respect causality 
and they are unstable in the sense that 
 linear perturbations 
of their global equilibrium states become unbounded
on a very short time scale. These important conclusions have been derived in  \cite{His2, His3} by analyzing linearized perturbations 
in the form of exponentially plane waves  
 off a globally homogeneous equilibrium state propagating on a Minkowski spacetime.
 As long as one of the 
 coefficients  $(\zeta, \eta, k, \sigma)$  in (\ref{66})
 is different than zero, there exist  transversal and longitudinal exponentially growing modes.  Based on this property, other physically acceptable solutions of the perturbations equations exhibit also this type of  instability
(for more details, consult \cite{His2}).\\
 
The
  results in 
   \cite{His2, His3}  revealed
  the following 
 highly counter-intuitive property: an equilibrium state appear to be stable when it is observed from a particular
 frame, but becomes unstable
 when is observed from a frame related to the first one by a Lorentz boost.
 This behavior can be seen but analyzing the behavior of 
 the perturbing plane waves modes constructed in 
 \cite{His2}.  
 For instance,
 whenever the equilibrium state is observed from the comoving frame, 
 then for the Eckart theory 
 (or more generally for any theory within the 
 Hiscock-Lindblom class) 
 where $k\neq 0$,  
   these modes
 contain exponentially growing modes while
 for the Landau-Lifshitz theory (where 
 $k=0$)  
 these plane waves
 contains only decaying modes. However, when the state
 is viewed from a Lorentz frame
 where the state
 is in motion, 
 then for both theories i.e. 
 Eckart and also the
 Landau-Lifshitz theory
 (or more generally 
 for all of first order theories) 
  contain exponentially growing modes and thus exhibit instability.
 Here one sees that within 
 the Landau-Lifshitz theory 
 the equilibrium state is stable when is observed from the comoving with the 
 state frame, but is unstable when is observed from  any other boosted frame.\\
 
 This counter intuitive 
 behavior has triggered a sizable  amount of research activity (see for example
 \cite {RP1, Per,
 GAV2, 
 Nor2, GAV6}) and these efforts lead to powerful statements linking causality to stability of equilibrium states.
 One of the strongest results describing this interplay is 
 Gavassino's criterion \cite{GAV2} which asserts that if for a theory 
 causality holds, 
 then stability of equilibrium sates is a Lorentz-invariance property. In more intuitive terms,
 the criterion asserts that
 if for a causal theory one is able to prove stability of an 
 equilibrium state relative to one reference frame, then there cannot be any growing Fourier mode in any other boosted frame.
  A slightly different statement  has proven also  in \cite{Nor2} which 
 restricts slightly however the nature of the field equations.\\ 
In view of these developments, an interesting interpretation regarding the origin of the instability 
in the Hiscock-Lindblom class seen in 
\cite{His2, His3},
has been put forward in ref. \cite{GAV6}. It was shown in that reference 
that for these theories the total entropy 
as a function of the accessible states fail to have
upper bound a situation which is in sharp contrast to what occurs for instance for the case
of the Israel-Stewart transient thermodynamics where the total entropy exhibits an absolute maximum
(for the stability properties of that theory consult 
\cite{His1}, \cite{His3}, \cite{Olson}). Finally very recently, 
a connection between causality and thermodynamical stability has been discussed in \cite{GAV3}.
It was shown in that work for any theory
that is thermodynamically stable i.e. the total entropy is maximized at equilibrium, it is also causal at least close to equilibrium, a conclusion that indicates clearly that 
causality is tied up 
thermodynamic stability (and not hydrodynamic stability) that is commonly  believed
(see illuminating  discussion in  \cite{GAV3}).\\
%The goal of this letter is to finally explain simply the relationship between causality and stability. We prove, with a geometrical argument, that if a theory is thermodynamically stable, namely if the entropy is maximised at equilibrium (see Gavassino [10]), it is also causal, close to equilibrium1. We show that the key to understand this result from a physical perspective is the underlying relationship between entropy and information. Furthermore, we explain why causality alone does not imply stability (see e.g. [11, 12]), but one needs at least to prove stability in a particular reference frame (in agreement with [7]).

In view of this perplexing
state of affairs 
centered on
  causality,  stability and the Hiscock-Lindblom class of first order theories,
  below we   
discuss briefly a few properties 
 of equilibrium states perturbed by states 
 satisfying the relativistic (LTE) postulate.
%In that regard, at first we demonstrate the expectable property that 
 %equilibrium states of a simple fluid propagating on a Minkowski spacetime
%are states compatible with the (LTE) postulate.
For this, let again the class of states satisfying the relativistic (LTE) postulate
assuming here that the fields $T^{\mu\nu}$ and $J^{\mu}$ 
are defined on a Minkowski spacetime
restricted so that the pseudo-angle 
$\epsilon$ 
in
(\ref{PSA}) satisfies
$\epsilon<<1$ 
everywhere
within the fluid region.
Let a four velocity 
$u_{o}^{\mu}$ is chosen within the cone of opening angle 
$\epsilon<<1$ 
and let  the fields 
in the expansion (\ref{OD}, \ref{OD_11})
satisfy
\begin{equation}
h^{\mu}=n^{\mu}= \pi=\pi^{\mu\nu}=0,
\label{P1}
\end{equation}
while $(\rho_{o}, n_{o}, P_{o})$ are chosen to be homogeneous and isotropic
and 
relative to a global
inertial coordinates $(t, x, y, z)$ 
 the velocity field
$u_{o}^{\mu}$ 
takes  the form
$u_{o}^{\mu}=\delta^{\mu}{}_{t}$ i.e. the state is at rest relative to this global rest frame.
The so defined state satisfies  the
relativistic (LTE) postulate but it also is an equilibrium state\footnote{Here after
thermodynamical variables describing this equilibrium state are denoted by subscript $(o)$.
In particularly, for emphasis we write 
$(T_{o}^{\mu\nu}, J_{o}^{\mu})$ instead of 
writing the correct expressions $(T^{\mu\nu}, J^{\mu})$.} since 
the entropy current $S^{\mu}$ as
a consequence of 
(\ref{P1}) satisfies $\nabla_{\mu}S^{\mu}=0$
while its energy momentum tensor $T_{o}^{\mu\nu}$
 and particle current $J_{o}^{\mu}$ satisfy
\begin{equation}
\nabla_{\mu}T_{o}^{\mu\nu}=\nabla_{\mu}[(\rho_{o}+P_{o})u_{o}^{\mu}u_{o}^{\nu}+P_{o}g^{\mu\nu}]=0,
 \quad \nabla_{\mu} J_{o}^{\mu}=\nabla_{\mu}(n_{o}u_{o}^{\mu})=0.
\label{P22}
\end{equation}
We perturb the equilibrium state in (\ref{P1}, \ref{P22})
by 
a state satisfying the (LTE) postulate denoted  by $(u^{\mu}, s(\rho, n))$.\\
 Following  the notation in
\cite{His2},
 the (Eulerian) perturbations of the thermodynamical variables 
 are denoted by 
$\delta \rho, \delta n, \delta u^{\mu},
\delta \tau, $ etc and represent
the difference between the value of the non equilibrium variable and the 
corresponding equilibrium one evaluated at the same spacetime point. It is not difficult to show
 that the equations for these 
perturbations are obtained by linearizing 
(\ref{FOD}-\ref{5}) around a background equilibrium state are given by (see also \cite{His2})

\begin{equation}
\nabla_{\mu}\delta T^{\mu\nu}=0,\quad 
\nabla_{\mu}\delta J^{\mu}=0,\quad
\delta h^{\mu}=-kT\Delta^{\mu\nu}[\nabla_{\nu} (\frac {\delta T}{T})+u^\lambda \nabla_{\lambda} \delta u_{\nu}+
\delta u^{\lambda} \nabla_{\lambda} u_{\nu}],\quad  \delta \nu^{\mu}=-\sigma T^{2} \Delta^{\mu\nu}\nabla_{\nu}\delta \Theta 
\label{P3}
\end{equation}
\begin{equation}
 \delta \pi^{\mu\nu}=-2\eta <\nabla ^ {\mu}\delta u^{\nu}+\delta u^{\mu} u^{\lambda}\nabla_{\lambda}u^{\nu}>,
 \quad \delta \pi=-\zeta \nabla_{\mu} \delta u^{\mu},
\label{P4}
\end{equation} 
 with the perturbations $\delta T^{\mu\nu}$ and $\delta J^{\mu}$ defined by

\begin{equation}
\delta T^{\mu\nu}=(\rho+P)(\delta u^{\mu}u^{\nu}+u^{\mu} \delta u^{\nu}) +\delta \rho u^{\mu}u^{\nu}+(\delta P+\delta \pi)
\Delta^{\mu\nu}+u^{\mu}\delta h^{\nu}+u^{\nu}\delta h^{\mu}+\delta \pi^{\mu\nu},
\label{P5}
\end{equation}
\begin{equation}
\delta J^{\mu}=\delta n u^{\mu}+n\delta u^{\mu}+\delta n^{\mu}.
\label{P6}
\end{equation}

and in these equations and here after, variables without the prefix $\delta$, denote equilibrium values\footnote{For typographical convenience we have written $(T, u^{\mu}, \Delta^{\mu\nu}, P,....)$ instead of $(T_{o}, u_{o}^{\mu}, 
\Delta_{o}^{\mu\nu}, P_{o},....)$.}.
Notice also that as a consequence of the constraints $u^{\mu}u_{\mu}=-1$, $h^{\mu}u_{\mu}=0$ etc, the perturbations
variables  are subject to the constraints (see also \cite{His2}):
\begin{equation}
u^{\mu}\delta u_{\mu}=u^{\mu}\delta h_{\mu}=u^{\mu}\delta n_{\mu}=u^{\mu}\delta \pi_{\mu\nu}=0
\label{P6}
\end{equation}

Since in the derivation of equations
(\ref{P3}-\ref{P6}),
are actually independent
of whether the perturbing state is an arbitrary state or a state compatible with the (LTE) postulate,
the results in \cite{His2}
holds true for our problem as well. Thus for small (Eulerian) perturbations
of exponential plane waves  of the form

\begin{equation}
\delta Q=\delta Q_{0}e^{ik_{x}x+\Gamma t}
 \label{P7}
\end{equation}
 propagating along the $x$-axis with constant ''frequency'' $\Gamma$, the results
 in
\cite{His2} 
 show 
 %and set \cite{His2}, and setting
%$$\delta Y^{B}:=(\delta \rho, \delta n, \delta u^{x}, \delta \pi, \delta h^{x}, \delta n^{x}, \delta \pi^{xx}, \delta u^{y}, \delta h^{y},\delta \pi^{xy},\delta u^{z}, \delta h^{z}, \delta \pi^{xz}, \delta n^{z}, \delta \pi^{yz}, \delta n^{y})$$,
that for a non vanishing $k_{x}$, exist exponentially growing (and decreasing) transverse modes
with real ''frequencies'' $\Gamma_{\pm}$  given by
\begin{equation}
2kT\Gamma_{\pm}=(\rho+P) \pm[(\rho+P)^{2}+4\zeta k T k_{x}^{2}]^{\frac {1}{2}} 
\label{P8}
\end{equation}
while for the case of the  Landau-Lifshitz theory there exist decreasing transverse modes with 
\begin{equation}
\Gamma=- \frac {\eta  k_{x}^{2}}{\rho+P},
\label{P9}
\end{equation}
and since this $\Gamma$ is purely real and negative, it follows that the Landau-Lifshitz 
theory escapes 
the instability that is manifest in the Eckart (or any other first order theory subject to $k>0$).
However, as was shown in 
 \cite{His2},  \cite{His3} this property disappears once
 the equilibrium state is viewed from a Lorentz frame
 where the equilibrium state is in motion.\\
 The results expressed in 
(\ref{P8}, \ref{P9}) hold 
also for the equilibrium states perturbed by states satisfying the relativistic (LTE) postulate
and thus these equilibrium states exhibit the same instabilities as the 
equilibrium states within the full class of first order theories.\\

Suppose however, we consider
another state satisfying the relativistic (LTE) postulated
specified
by $(\hat u^{\mu}, s(\hat \rho, \hat n))$,
which is initially
close to 
the equilibrium state in (\ref{P22}).
We consider again
the linear perturbations
denoted here after by a prime i.e.
$\delta' \rho, \delta' n, \delta'u^{\mu},
\delta'\tau, $ etc induced by 
$(\hat u^{\mu}, s(\hat \rho, \hat n))$
on the 
equilibrium state in (\ref{P22}).
Clearly  these new perturbations satisfy 
again (\ref{P3}-\ref{P6})
(obtained by linearizing the system 
(\ref{FOD}-\ref{5})).\\
As we have seen in this section although,
the states  
 $(u^{\mu}, s(\rho, n))$
and 
 $(\hat u^{\mu}, s(\hat \rho, \hat n))$
are in fact equivalent
and formulas
 (\ref{lema1_02}-\ref{lema3_08}),
shows that
equivalence, below
we show that 
the perturbations induced by 
 $(u^{\mu}, s(\rho, n))$
and 
 $(\hat u^{\mu}, s(\hat \rho, \hat n))$
on the 
background 
equilibrium state in  (\ref{P22})
fail to be equivalent.\\
%The primed and unprimed versions of 
%(\ref{P1}, \ref{P2}) constitute a complicated system of linear equations and the 
%behavior of their solutions are described by
%the Hiscock-Linblom solutions in (\cite{His2}) (see also the approach of Kovtun in 
%(\cite{Kov2}).)
%As longs as the perturbing states  both satisfying the relativistic (LTE) postulate,
%then the two systems of perturbation equations i.e. the unprimed and primed versions of 
%(\ref{P3}-\ref{P6})
%are  equivalent sets i.e. solutions of one can transformed to solutions of the other and vice versa.
%In fact the situations is analogues to what occurs for the equations describing first order deviations from a state of 
 %''local thermodynamical equilibrium'' specified say by $(\hat u^{\mu}, s(\rho, n))$. 
Indeed starting from the relation
$$\hat u^{\mu}-u^{\mu}=\hat \epsilon^{\mu}+O(\hat \epsilon)^{2}$$ 
  the following relations hold between the 
 primed and unprimed perturbations: 
 
 \begin{equation}
\delta u^{\mu}:= u^{\mu}- u_{o}^{\mu}=\hat u^{\mu}- u_{o}^{\mu}+u^{\mu}-\hat u^{\mu}=\delta'u^{\mu}-\hat \epsilon^{\mu}+
O(\hat \epsilon)^{2}
\label{P10}
\end{equation}
\begin{equation}
\delta h^{\mu}= h^{\mu}(u)=\delta' h^{\mu} +(\rho+P)\hat \epsilon ^{\mu}+O(\hat \epsilon)^{2},
\quad
\delta n^{\mu}=\delta 'n^{\mu}+n\hat \epsilon ^{\mu}+O(\hat \epsilon)^{2}
\label{P11}
\end{equation}

\begin{equation}
 \delta \rho=\delta'  \rho +O(\hat \epsilon)^{2},\quad
 \delta n=\delta' n +O(\hat \epsilon)^{2},\quad  
 \quad  \delta P=\delta' P +O(\hat \epsilon)^{2} 
 \quad  \delta \pi= \delta' \pi+O(\hat \epsilon)^{2},\quad 
 \delta \pi^{\mu\nu}=
  \delta' \pi^{\mu\nu}+O(\hat \epsilon)^{2} 
\label{P12}
\end{equation}

Moreover it can be easily seen that if the unprimed perturbations
satisfy equations (\ref{P3}-\ref{P6}). This means that
the perturbations induced
on the background equilibrium state
 by 
 $(u^{\mu}, s(\rho, n))$
and 
 $(\hat u^{\mu}, s(\hat \rho, \hat n))$
states are in fact related
by formulas 
(\ref{P10}-\ref{P12}) that are 
analogous to (\ref{lema1_02}-\ref{lema3_08}).
Since however, these relations hold for arbitrary solutions of the perturbations equations,
they also hold for the 
the 
of exponential plane waves  
solutions described 
 in (\ref{P7}) i.e.
 propagating along the $x$-axis and of constant ''frequency'' $\Gamma$, as in
\cite{His2}.
However, here we lead into an impasse. If for instance we choose
 $(u^{\mu}, s(\rho, n))$ to be 
described by  an Eckart state ( or more generally an arbitrary
 first order state with $k>0$) and 
 $(\hat u^{\mu}, s(\hat \rho, \hat n))$
say a Landau-Lifshitz state (thus
 $k=0$), then formulas 
 (\ref{P10}-\ref{P12})
 describe an incompatibility.
 One perturbation is exponentially growing while the other is exponentially decreasing.
 This in turn implies that the equivalence between 
 the two states 
 $(u^{\mu}, s(\rho, n))$
and 
 $(\hat u^{\mu}, s(\hat \rho, \hat n))$ 
 that satisfy the (LTE) postulate
 as expressed by  
 formulas
 (\ref{lema1_02}-\ref{lema3_08}),
 breaks down for the first order perturbations of an equilibrium state.
 To put it differently, if the states 
  $(u^{\mu}, s(\rho, n))$
and 
 $(\hat u^{\mu}, s(\hat \rho, \hat n))$ 
were truly equivalent  one would have expected
that the induced perturbations on a background equilibrium state of the theory to in fact identical
solutions.\\
 Interestingly this kind of equivalence holds for the Israel-Stewart theory. If one
  replaces the Hiscock-Libdblom class of first order theories by the 
Israel-Stewart theory and the
$(u^{\mu}, s(\rho, n))$,
 $(\hat u^{\mu}, s(\hat \rho, \hat n))$ 
states by the 
Israel-Stewart theory expressed in the Eckart frame respectively 
Landau-Lifshitz frame, one would find that the resulting induced perturbations are indeed equivalent
and this equivalence has been demonstrated\footnote{Our thanks to an 
  anonymous referee who pointed out to us ref.\cite{GAV1}
 and suggested to probe the interconnection between 
  Hiscock-Libdblom class of first order theories and the Israel-Stewart theory.}
 in ref. \cite{GAV1} (we shall return to this point at the end of the section $(V)$). Here
 we only mentioned that this pronounced difference between the 
 two theories can 
 be traced in thermodynamical reasoning.
Gavasinno, Antonelli and Haskell in \cite{GAV6} have shown that the total entropy 
between the two theories behaves differently.
Sable point behavior for the 
 Hiscock-Libdblom class of first order theories
versus an absolute maximum 
for the 
 Israel-Stewart theory. We shall discuss this point further at the
 end of the next section.\\
 The so far analysis  demonstrates that states
 satisfying 
the relativistic (LTE) postulate  with the 
  Hiscock-Libdblom class of first order theories
  exhibit the same pathologies as arbitrary states
  within this class and thus their utility is very limited.\\

We shall  leave this section, by briefly discussing 
the connection between states obeying the relativistic (LTE) postulate 
%and states belonging%
 and the 
 (BDNK) theory. 
 As we have already mentioned in 
  section $(II)$  
 the (BDNK) theory is branded as a first order theory\footnote{It ought to be stressed that while in the Hiscock-Lindblom terminology, a first order theory is a theory where the entropy  flux $S^{\mu}$ receives contributions only
from  first order deviations from a
fictitious ''local thermodynamical equilibrium'' 
state, the term ''first order theories'' 
within the 
(BDNK) formalism 
has a very different meaning as it will become clear further ahead. A suitably defined series is approximated only by the first term.}
and it
has some remarkable properties.
It was  shown in  (\cite{Kov1}, \cite{Kov2}, 
\cite{Nor1}, \cite{Nor2}) that under suitable choice of frame, the theory respects 
causality 
and admits
stable equilibrium states
and moreover in 
\cite{Nor2}
it was  shown that 
the Cauchy problem
for the theory is locally well posed, and strongly hyperbolic
and this
 locally well posedness and strong hyperbolicity remain intact even
when the fluid is dynamically coupled 
to Einstein's equations ( see ref \cite{Nor2} for details and references).\\

For a simple, electrically neutral, fluid
the dynamical equations for this theory are  
$\nabla_{\mu}T^{\mu\nu}=\nabla_{\mu}J^{\mu}=0$
and fluid states 
 are specified by assigning a ''hydrodynamical frame'' or simply a frame\footnote{Here we are warning the reader of a potential confusion arising from the use of the term ''frame''. While in this work the term ''frame''
signifies an orthonormal tetrad (or
often a collections of orthonormal tetrads
defined along an integral curve (or on all integral curves)) 
of a smooth velocity field $u^{\mu}$
and the term ''change of frame'' is
 a transition to a new tetrad induced by the action of a point wise Lorentz transformation (or a family of such point wise transformations), within
 the (BDNK) formalism, 
 a ''hydrodynamical frame'' or simply ''frame'' signifies a specification of a triplet
 $(T, \mu, u^{\mu})$ and a change of the ''hydrodynamical frame''
(or a field redefinition), is a passage to a new 
 ''hydrodynamical frame''
i.e. to a new set of hydrodynamical variables $(\hat T, \hat \mu, \hat u^{\mu})$. This change of frame advocated in the
(BDNK) formalism
is in general different than the change of frame induced by the action of a
local Lorentz transformation in our terminology, although the two are related as we shall see further ahead 
(in reference (\cite{Nor2}) a complete set of references are compiled where 
the reader can find the 
diverse meaning assigned to the term frames and change of frames).}
which signifies a specification of a fluid four velocity
 $u^{\mu}$ a temperature $T$ and a chemical potential $\mu$ (both of them measured by the 
 $u^{\mu}$-observer). The theory 
 targets states 
 that are near equilibrium, although
a precise definition
of what exactly constitutes 
a state near thermal equilibrium
has not been addressed in the so far development of the theory\footnote{ An attempt to define precisely the
term equilibrium and states near equilibrium
is pursued in refs
\cite{Nor1}, \cite{Nor2}
via the employment of relativistic kinetic theory. 
Although a
Chapman-Enskog expansion, 
 or a Grad like or any other expansion yield
fluid  states that capture the spirit of the relativistic (LTE) postulate, still we feel that
an independent definition of states compatible with a relativistic (LTE) postulate
build entirely within the field of Relativistic hydrodynamics viewed as a discipline 
 in its own right,
 is worth having (and this is precisely the main purpose of the present work).}.
Within this theory,  the fields
$(T^{\mu\nu}, J^{\mu})$ depend upon $(u^{\mu}, T, {\mu})$
and the theory
utilizes (or better adapts) the 
gradient expansion technique from quantum field theory to 
the hydrodynamical regime (for an introduction to this 
 technique within the hydrodynamical regime, 
consult (\cite{Gex1}, \cite{Gex2}) and references therein).
The fields $(T^{\mu\nu}, J^{\mu})$ are expanded according to

\begin{equation}
 T^{\mu \nu}  = \mathcal{E} u^\mu u^\nu + \mathcal{P}\Delta^{\mu \nu}(u) + \left(\mathcal{Q}^\mu u^\nu + \mathcal{Q}^\nu u^\mu\right) + \mathcal{T}^{\mu \nu}\\
 \label{EXP0}
\end{equation}

\begin{equation}
J^{\mu} = \mathcal{N}u^\mu + \mathcal{J}^\mu
  \label{EXP00}
\end{equation}

which are the same expansions as in 
(\ref{OD}, \ref{OD_11}) except that new symbols\footnote{The usage of 
$(\mathcal{E},\mathcal{P},\mathcal{Q}^\mu,
\mathcal{T}^{\mu \nu},\mathcal{N},\mathcal{J}^\mu)$ is the standard notation
for the practioners of the (BDNK) formalism.} $(\mathcal{E},\mathcal{P},\mathcal{Q}^\mu,
\mathcal{T}^{\mu \nu},\mathcal{N},\mathcal{J}^\mu)$
are used instead of 
$(\rho, P, h^{\mu}, \tau^{\mu\nu}, n, n^{\mu})$ entering in 
(\ref{OD}, \ref{OD_11}).
However, within the (BDNK) formalism,
it is postulated that the fields
$(\mathcal{E},\mathcal{P},\mathcal{Q}^\mu,
\mathcal{T}^{\mu \nu},\mathcal{N},\mathcal{J}^\mu)$
admit a (convergent)  derivative series expansion 
formed from the derivatives of  $(T, u^{\mu},\mu) $.
The series employs
 the scalars $(T, \mu)$, % and no transverse vectors, and no transverse traceless 2-tensors.%, 
 their derivatives along the flow of $u^{\mu}$
 i.e. $u^{\mu} \nabla_{\mu} T :=\dot T$,   
$\nabla_{\mu} u^{\mu}$ and $u^{\alpha} \nabla_{\alpha} \mu:=\dot \mu $,
 utilizes also
 the transverse vectors 
$\Delta^{\mu\nu}\nabla_{\nu} T$, $a^{\mu} :=u^{\nu}\nabla_{\nu} u^{\mu}$,
$\Delta^{\mu\nu}\nabla_{\nu} \mu$, and the transverse second rank, traceless symmetric tensor, $\sigma_{\mu\nu}$
and in a standard notation, 
the series expansions 
have the form
(for details consult (\cite{Kov1}, \cite{Kov2}, 
\cite{Nor1}, \cite{Nor2}):

\begin{equation}
\mathcal{E}=\epsilon+\epsilon_{1} \frac {\dot T}{T}+\epsilon_{2}\nabla_{\mu} u^{\mu}+\epsilon_{3}u^{\mu}\nabla_{\mu} (\frac {\mu}{T})+
O(\partial^{2})
\label{EXP1}
\end{equation}

\begin{equation}
\mathcal{P}=p+\pi_{1} \frac {\dot T}{T}+\pi_{2}\nabla_{\mu} u^{\mu}+\pi_{3}u^{\mu}\nabla_{\mu} (\frac {\mu}{T})+
O(\partial^{2})
\label{EXP2}
\end{equation}

\begin{equation}
\mathcal{Q}^{\mu}=\theta_{1} \dot u^{\mu}+\frac {\theta_{2}}{T} \Delta^{\mu\nu}\nabla_{\nu} T+\theta_{3}
\Delta^{\mu\nu} \nabla_{\nu} (\frac {\mu}{T})+
O(\partial^{2})
\label{EXP3}
\end{equation}

\begin{equation}
\mathcal{T}^{\mu\nu}=-\eta \sigma^{\mu\nu}+
O(\partial^{2})
\label{EXP4}
\end{equation}

\begin{equation}
\mathcal{N}=n+\nu_{1} \frac {\dot T}{T}+\nu_{2}\nabla_{\mu} u^{\mu}+\nu_{3}u^{\mu}\nabla_{\mu} (\frac {\mu}{T})+
O(\partial^{2})
\label{EXP5}
\end{equation}

\begin{equation}
\mathcal{J}^{\mu}=\gamma_{1} {\dot  u}^{\mu}+\frac{ \gamma_{2}}{T} \Delta^{\mu\nu} \nabla_{\nu} \frac {\mu}{T}+\gamma_{3}\Delta ^{\mu\nu}\nabla_{\nu} 
\frac {\mu}{T}+
O(\partial^{2})
\label{EXP6}
\end{equation}

where $O(\partial^{2})$ (or more generally 
 $O(\partial^{k})$)
signifies terms of second order (or  $k$ order) derivatives in the variables
$(T, u^{\mu},\mu) $,
while 
 $(\epsilon_{i}, \pi_{i},\theta_{i}, \nu_{i})$,~ $(\gamma_{i}, i\in (1,2,3))$ and $\eta$
are considered to be transport coefficients depending upon  $(T, \mu)$. 
%Using the freedom in the choice of the hydrodynamical frame, it turns out that only 
%three of these $(16) $ transport coefficients are really independent and this reduction is due to the liberty in the choice of the 
%hydrodynamical frame  $(T, u^{\mu},\mu) $.
Moreover, it is postulated that the basic fields 
$(T(x), u^{\mu}(x),\mu(x))$ are non unique and can be transformed to new fields
$(T'(x), u'^{\mu}(x),\mu'(x))$ according to:

\begin{equation}
(T(x), u^{\mu}(x),\mu(x)) \to (T'(x), u'^{\mu}(x),\mu'(x))=(T(x)+\delta T(x), u^{\mu}(x)+\delta u^{\mu}(x), 
\mu(x)+\delta \mu(x)))
\label{EXP7}
\end{equation}

where the variations 
$(\delta T(x), \delta u^{\mu}(x), \delta \mu (x))$
are defined via:
\begin{equation}
\delta T=\alpha_{1} \frac {\dot T}{T}+\alpha_{2}\nabla_{\mu} u^{\mu}+\alpha_{3}u^{\mu}\nabla_{\mu} (\frac {\mu}{T})+
O(\partial^{2})
\label{EXP8}
\end{equation}

\begin{equation}
\delta u^{\mu}=b_{1}{\dot u^{\mu}}+b_{2}\Delta^{\mu\nu}\nabla_{\nu} T
+b_{3}\Delta^{\mu\nu}\nabla_{\mu} (\frac {\mu}{T})+
O(\partial^{2})
\label{EXP9}
\end{equation}

\begin{equation}
\delta \mu=
c_{1} \frac {\dot T}{T}+c_{2}\nabla_{\mu} u^{\mu}+c_{3}u^{\mu}\nabla_{\mu} (\frac {\mu}{T})+
O(\partial^{2})
\label{EXP10}
\end{equation}
where $(a_{i}, b_{i}, c_{i}, i\in (1,2,3))$
are arbitrary functions of the $(T,\mu)$
(for a discussion leading to  the above transformation see
 \cite{Kov1}, \cite{Kov2}).
The transformation in 
(\ref{EXP7}-\ref{EXP10})
describes a change of the hydrodynamical frame
and plays an important role within the theory.\\

We shall not discuss any further properties of this theory, but 
for the rest of this section, we shall link 
 states obeying the relativistic (LTE) postulate 
 to the (BDNK) theory and for this let us 
 begin by 
 choosing a 
 state compatible with 
the relativistic (LTE) postulate
within however the (BDNK) theory.
Accordingly we choose $T^{\mu\nu}$ and $J^{\mu}$ 
 such that
the angle $\epsilon$ formed by 
 $u^{\mu}_{E}$ and $u^{\mu}_{N}$
(see (\ref{PSA})) satisfies 
 $\epsilon<<1$ everywhere within the the fluid region
 and choose 
a velocity field $u^{\mu}$
within the cone of the angle $\epsilon <<1$.
We denote by 
 $(u^{\mu}, s(\rho, n)) $ be the state of
''local thermodynamical equilibrium'' and 
let 
$T(u)$ the  local temperature and $\mu(u)$ 
the
 local chemical potential, defined from derivatives of $s(\rho(u), n(u))$
(see (\ref{OD_22}) and remember that $\Theta(u)=\frac {\mu(u)}{T(u)}$). 
Thus any state satisfying the relativistic (LTE) postulate leads to a 
hydrodynamical frame 
(in the (BDNK)-sense)
specified by 
$(T(u), u^{\mu}, \mu(u))$. \\

 Under an admissible frame change
  described by 
 \begin{equation}
u^{\mu} \to \hat u^{\mu}=u^{\mu}+\hat{\epsilon}^{\mu}+O(\hat{\epsilon})^{2},\quad  \hat{\epsilon}^{\mu} \leq O_1.
\label{FCN}
\end{equation} 
formulas (\ref{lema1_02}-\ref{lema3_08}),
imply 
\begin{equation}
T':=T(\hat u)=T(u)+\hat \epsilon O_{1},\quad  \mu':=\mu'(\hat u)=\mu(u)+\hat \epsilon O_{1},\quad u'^{\mu}:=\hat u^{\mu}=u^{\mu}+\hat \epsilon^{\mu}+ (\hat \epsilon)^{2}
\label{FCN11}
\end{equation}
where 
$(T(\hat u), \mu (\hat u))$  are the local temperatures and local chemical potentials measured by the $\hat u^{\mu}$-observer.
Thus any 
 frame change as 
 in (\ref{FCN}), induces a particular field redefinition
 described in 
(\ref{FCN11}) and this field redefinition can be seen as a particular case of the field redefinition in
(\ref{EXP7}-\ref{EXP10}). Indeed it can be
generated by choosing
 $\alpha_{i}=c_{i}=0, i \in (1,2,3)$, taking $b_{1}=b_{2}= b_{3}=b << 1$ 
 in (\ref{EXP8}-\ref{EXP10}),
 and identify
 $\hat \epsilon^{\mu}$ in (\ref{FCN}) via
 
 \begin{equation}
 \hat \epsilon^{\mu}:=
 b[{\dot u^{\mu}}+\Delta^{\mu\nu}\nabla_{\nu} T
+\Delta^{\mu\nu}\nabla_{\mu} (\frac {\mu}{T})+
O(\partial ^{2})].
\label{FCN12}
\end{equation}
 Under the field redefinition in 
(\ref{FCN11}), the series expansion 
in
(\ref{EXP1}
-\ref{EXP6}) remain form invariant, since terms like $\nabla_{\mu} \hat \epsilon_{\nu}$ and 
$\nabla_{\mu} \hat \epsilon^{\mu}$ are considered as second order in the deviation from the state 
state of  ''local thermodynamical equilibrium'' and the contributions of terms like $u^{\mu}\hat \epsilon ^{\nu}\nabla_{\nu} T$,
$\hat \epsilon^{\mu}\hat \epsilon ^{\nu}\nabla_{\nu} T$ etc are neglected in comparison to terms
$u^{\mu}u ^{\nu}\nabla_{\nu} T$,  $u^{\mu} u^{\nu}\nabla_{\nu} T$ etc.
The invariance of the series
in (\ref{EXP1}-\ref{EXP6}) implies that the dynamical equations satisfied by 
$(T, \mu, u^{\mu})$
remain form invariant under the transformation of frame induced by 
(\ref{FCN}) as long as second order deviations from a ''local 
thermodynamical equilibriumÓ state
are neglected. Thus
within the (BNTK) formalism
states compatible to 
the relativistic (LTE) postulate are equivalent
provided second order deviations from 
from a ''local 
thermodynamical equilibrium'' state
are thrown away.\\
However ought to be stressed that the real power of the 
the (BDNK) formalism lies in the freedom of  frame change described 
in (\ref{EXP7}-\ref{EXP10}). This liberty, allows to select particular hydrodynamical 
frames
where the second order equations 
$\nabla_{\mu}T^{\mu\nu}=\nabla_{\mu}J^{\mu}=0$ for the basic variables
$(u^{\mu}, T, {\mu})$
(or an equivalent set) exhibit causality 
and stability of equilibrium states (for details consult
(\cite{Kov1}, \cite{Kov2}, 
\cite{Nor1}, \cite{Nor2}).
Thus within the 
(BDNK) framework, states satisfying 
the relativistic (LTE) postulate are becoming relevant. \\

\section{Transient thermodynamics and (LTE) states }
 
In this section, we discuss 
 states satisfying the relativistic (LTE) postulate
   within the Israel-Stewart \cite{Isr1, Isr2} transient thermodynamics.
 We ought to bear  in mind however, that 
transient thermodynamics, by design deals
 with states that are near equilibrium
 (or in our terminology with states
  satisfying the relativistic (LTE) postulate\footnote{
  As we have already mentioned, the definition of states  compatible with
  the relativistic (LTE) postulate proposed in this work, has been motivated by the properties of states satisfying the equations of transient thermodynamics.
  Also we mention here that a state near equilibrium it is not necessary a state satisfying
  the relativistic (LTE) postulate. The four velocity that identifies the former does not necessarily leads to 
  a state where ''local thermodynamical equilibrium'' prevails.}).
 It is worth remembering this restriction since
 as pointed in 
   \cite{His4} for states with   
   large deviations
  from equilibrium the theory it is not always well behaved.
 In that regard, is worth remembering 
 some recent results 
  derived  in ref. 
  \cite{BEM} where 
  sufficient conditions, 
  in the form of algebraic inequalities
  derived that ensure   
  that the theory is causal away from equilibrium
  and a theorem established on the  local existence and uniqueness of solutions of the theory
  and these results raise the confidence on the theory. \\
  With this comments in mind, and since
 the 
structure of the theory is discussed in the original articles,
 by Israel \cite{Isr1} and Israel and Stewart \cite{Isr2}
 below, 
 we shall only provide
  a brief overview of the principles of transient thermodynamics and 
  shall offer a few
  comments regarding the equivalence of the phenomenological  equations as expressed relative to distinct frames.\\

We begin by considering a 
 fluid state compatible with the relativistic (LTE) postulate described by
$(J^{\mu}, T^{\mu\nu}, S^{\mu} )$
obeying
(\ref{BE}, \ref{EL}) 
with the entropy flux $S^{\mu}$ to be defined shortly.
We choose a four velocity $u^{\mu}$ defined within the fluid region
and lying within the familiar cone of opening angle 
$\epsilon<<1$ and let
$s(\rho, n)$ be the  corresponding equilibrium equation of state.
To this 
$(u^{\mu}, s(\rho, n)),$ 
we attach 
the ''local thermodynamical equilibrium'' state
$(S^{\mu}_{0}, T^{\mu\nu}_{0}, J^{\mu}_{0})$
in the manner discussed 
in section $(III)$ (see discussion leading 
to (\ref{33})). The velocity  $u^{\mu}$
and the expansions 
in (\ref{OD},\ref{OD_11}), 
define the fields 

\begin{equation}
(n( u),  \rho(u),  P(u),  h(u)^{\mu},  n(u)^{\mu}, \tau(u)^{\mu\nu})
\label{TRA_100}	
\end{equation}
 while 
 the thermal potential 
 $\Theta(u)$
and the local inverse temperature
$ \beta(u)=T(u)^{-1}$
measured by the $u$-observer
are defined by the fundamental Gibbs relation
(\ref{OD_22}).
The velocity
$u^{\mu}$
and the fields 
in (\ref{TRA_100})	
 satisfy the eqs of transient thermodynamics 
provided they obey:

\begin{widetext}
\begin{equation}\label{FOD1}
\nabla_{\mu}T^{\mu\nu}=\nabla_{\mu}[\rho(u)u^{\mu}u^{\nu}+P(u)\Delta(u) ^{\mu\nu}+h(u)^{\mu}u^{\nu}+h(u)^{\nu}u^{\mu}+\tau(u)^{\mu\nu}]=0,
\end{equation}
\end{widetext}

\begin{equation}
	\nabla_{\mu} J^{\mu}=\nabla_{\mu}[n(u)u^{\mu}+n(u)^{\mu}]=0,
\label{FOD_100}	
\end{equation}

and additional equations arising by imposing the second law.
The entropy flux $S^{\mu}$ is postulated to have the form (see \cite{Isr1, Isr2}  a discussion leading to this choice):

\begin{equation}
	S^\mu = P(u)\beta^\mu(u) - \Theta(u) J^\mu-
	\beta_\nu(u) T^{\mu \nu} - Q^\mu(\delta J^\alpha, \delta T^{\alpha \beta}, X^{\alpha \beta \gamma..}_{(i)})
\label{TT_04}
\end{equation}

where 
 $\beta_\lambda = \beta(u)u_\lambda = \frac{u_\lambda}{T(u)}$
 and $Q^\mu$ is a vector field 
that depends
quadratically upon 
the deviations  
$ (\delta J^\alpha, \delta T^{\alpha \beta})$  
 from the state\footnote{ The variables
 $X_{(i)}^{\mu\nu\lambda...}$,$i\in (1,2,....)$ appearing in $Q^{\mu}$,
 are additional variables needed to completely specify the non equilibrium state. In this work, and in fact within transient 
 thermodynamics, are assumed to be zero
 and thus will be disregarded here after.} 
  of a ''local thermodynamical equilibrium''
 specified  by $(u^{\mu}, s(\rho, n))$.\\
 For the following analysis, it is convenient to transform 
the right hand side of 
(\ref{TT_04}) in a form so that
 its relation to the 
entropy flux $S^{\mu}$ 
for first order theories, becomes apparent.
For this, starting from
$ s(u) = \beta(u)\left(\rho(u) + {P(u)}\right) - \Theta(u) n(u),$ and the fundamental Gibbs relation
 $ds = \beta d\rho - \Theta dn$, one gets their covariant versions
\begin{equation}
S^\mu_{0} = P\beta^\mu - \Theta J^\mu_{0} - \beta_\lambda T^{\lambda \mu}_{0},\quad
dS^\mu_{0} = -\Theta dJ^\mu_{0} - \beta_\lambda dT^{\lambda \mu}_{0}, 
\label{C1}
\end{equation}
where for their derivations we used: $S^{\mu}_{0}=su^{\mu}$, 
$\rho u^\mu = -u_\lambda T^{\lambda \mu}_{0}$ and $(J^{\mu}_{0}, T^{\mu\nu}_{0}, S^{\mu}_{0})$ are the fields
introduced in (\ref{33}).
%and  $b_\lambda = bu_\lambda = \frac{u_\lambda}{T}$. 
Eliminating  the contribution of 
$P(u)\beta^\mu$ from the right hand side of
(\ref{TT_04}) one gets
\begin{equation}\label{TT_05}
\begin{split}
	S^\mu & = S^{\mu}_{0}-\Theta (J^{\mu}-J^{\mu}_{0})-
	\beta_{\lambda}(T^{\lambda\mu}-T^{\lambda\mu}_{0})-Q^{\mu} \\
	& =S^{\mu}_{0} -\Theta \delta J^{\mu}-
	\beta_{\lambda}\delta T^{\lambda\mu}-Q^\mu(\delta J^\alpha, \delta T^{\alpha \beta}) \\
	& = su^{\mu}
-\Theta(u) n^{\mu}(u) +\beta(u) h^{\mu}(u) -Q^\mu(\delta J^\alpha, \delta T^{\alpha \beta}),
\end{split}	
\end{equation}
which shows that transient
thermodynamics is an extension of the Hiscock-Lindblom class first order theories
by incorporating in $S^{\mu}$
second order deviations from the state of ''local thermodynamical equilibrium''
manifesting themselves in the non vanishing
of the $Q^\mu(\delta J^\alpha, \delta T^{\alpha \beta})$ term.\\

Motivated from
the relativistic kinetic theory of diluted gases, 
Israel and Stewart \cite{Isr1, Isr2} proposed that 
$S^{\mu}$ (and thus $Q^{\mu}$), 
should be independent of the gradients of $J^{\mu}$ and $T^{\mu\nu}$ and 
should be quadratic in the 
deviations from the state of
''local thermodynamical equilibrium''.
Within the hydrodynamical approximation\footnote{The name derives itself  from the assumption that 
$S^{\mu}$
is specified only by the hydrodynamical variables
 $(J^{\mu}, T^{\mu\nu})$.},
 they proposed that for an
electrically neutral, simple fluid the term $Q^{\mu}(u)$
should have the form:

\begin{widetext}
\begin{equation}\label{FQ1}
Q^{\mu}(u)=
\frac {1}{2}u^{\mu}[\beta_{0}\pi ^{2}+\beta_{1}q^{\nu}q_{\nu}+\beta_{2}\pi^{\lambda\nu}\pi_{\lambda\nu}]-
\alpha_{0}\pi q^{\mu}-\alpha_{1}\pi^{\mu\nu}q_{\nu}+R^{\mu}(u),
\end{equation}
\end{widetext}

where $R^{\mu}(u)$ stands for
\begin{equation}
R^{\mu}(u)=\frac {1}{T(\rho+P)}\left[
\frac {1}{2}u^{\mu}h^{\nu}h_{\nu}+\tau^{\mu\nu}h_{\nu}\right],
\label{FR}
\end{equation}
and in above
$q^{\mu}$
stands for  the 
 invariant heat flux  defined in 
 (\ref{lema2_01}). The 
 coefficients 
 $\alpha_{j}$, $j \in (0, 1)$, $\beta_{i}$, 
$i \in (0, 1, 2)$,
are 
undetermined
 depending upon $(\rho, n)$
and  $(T, \rho ,P,  \pi, etc)$ are the local thermodynamical variables
measured by the $u^{\mu}$-observer (for simplicity of the representation we
 suppressed their implicit dependance upon $u^{\mu}$).\\
As we mentioned, the dynamical equations for transient thermodynamics,
are those in  (\ref{FOD1},\ref{FOD_100})	
augmented by the 
phenomenological equations
that follow by imposing the second law
on the  entropy flux $S^{\mu}$ in
 (\ref{TT_04})
and $Q^{\mu}(u)$ as in
(\ref{FQ1}, 
\ref{FR}). Their forms and 
the detail calculations leading to their derivations
can be found  in  \cite{Isr1,Isr2}.
Here we shall offer a few comments regarding
their behavior under a frame change.
For this, we consider another velocity
field $\hat u^{\mu}$ lying within  
the cone of the opening pseudo-angle 
$\epsilon<<1$ and let the frame change
\begin{equation}
u^{\mu}\to \hat u^\mu =u^{\mu}+\hat \epsilon ^{\mu}, \quad
	\hat \epsilon^{\mu} \leq \epsilon^{\mu}.
	\label{VRTR}
\end{equation}
For this frame change, we introduce 
the new fields
$n(\hat u),  \rho(\hat u),  P(\hat u),  h(\hat u)^{\mu},  n(\hat u)^{\mu}, \tau(\hat u)^{\mu\nu}$
defined according to
$$ n(\hat u)= n( u)+\hat \epsilon O_{1},\quad \rho(\hat u)= \rho( u)+\hat \epsilon O_{1},\quad etc,$$
and at first we show that 
these new fields  satisfy to an $O_{1}$
accuracy the conservation laws
(\ref{FOD1},
\ref{FOD_100}). Indeed a repetition of 
the calculation that lead us 
to (\ref{LMR011},\ref{LMR022}) 	
for the case of first order theories, yields to  the same conclusion.
To an  $O_{1}$ accuracy in the deviations from the state of ''local thermodynamical equilibrium''
the conservation laws (\ref{FOD1},
\ref{FOD_100}) remain valid.\\
However 
proving that the same property holds for the rest of the
equations that follow 
 by imposing the second law on  $S^{\mu}$
in (\ref{TT_05})
becomes more subtle. One may for instance, start from the equations
written say relative to the $u^{\mu}$ frame and subsequently 
via 
 (\ref{lema1_02}-\ref{lema3_08}), rewrite them relative to the $\hat u^{\mu}$ frame keeping only terms describing first order deviations. Unfortunately this procedure is long and time consuming, but there is an alternative method
that avoids this route. For this,
we return to  the entropy flux $S^{\mu}$ in (\ref{TT_05}) and examine its variation under the frame change in
(\ref{VRTR}). 
Denoting by $\delta$ the resulting variation, 
and noting that $\delta S^{\mu}=\delta J^{\mu}=\delta T^{\mu\nu}=0$, we  find 
that $\delta S^{\mu}$
in
(\ref{TT_05}) satisfies:

\begin{equation}\label{TT_05B}
\begin{split}
	0=\delta S^\mu = u^{\mu}\delta s + s\delta u^{\mu}-(J^{\mu}-J_{0}^{\mu})\delta \Theta-\Theta \delta 
	(J^{\mu}-J_{0}^{\mu}) \\
	-\delta \beta_{\lambda}(T^{\mu \lambda}-T_{0}^{\mu \lambda})-\beta_{\lambda}\delta(T^{\mu \lambda}-T_{0}^{\mu \lambda})
-\delta Q^{\mu}.	
\end{split}	
\end{equation}

Due to the identity $ s(u) = \beta(u)\left(\rho(u) + {P(u)}\right) - \Theta(u) n(u),$ 
one finds  that 
$s\delta u^{\mu}-\Theta \delta(J^{\mu}-J_{0}^{\mu})-	
\beta_{\lambda}\delta(T^{\mu \lambda}-T_{0}^{\mu \lambda})$
is vanishing to linear order in $\epsilon^{\mu}$
and thus 
(\ref{TT_05B}) implies
\begin{equation}\label{TR_06}
\delta Q^{\mu}=	-(J^{\mu}-J_{0}^{\mu})\delta \Theta	
	-(T^{\mu \lambda}-T_{0}^{\mu \lambda})\delta \beta_{\lambda} + \hat \epsilon O_{1}+\hat \epsilon O_{2}.
	\end{equation} 
Taking into account
that 
\begin{equation}
\begin{split}
	\delta \Theta & \equiv \Theta(\hat{u})-\Theta(u) = \epsilon O_{1},\\
\delta \beta^{\mu} & = \frac {\epsilon ^{\mu}}{T(u)} + u^{\mu} \delta\left(\frac {1}{T(u)}\right) = \frac {\epsilon^{\mu}}{T(u)}+\epsilon O_{1},\\[10pt]
\end{split}
\end{equation}     
results
\begin{equation}\label{TQF}
\delta Q^{\mu}=-(T^{\mu \lambda}-T_{0}^{\mu \lambda})\delta \beta_{\lambda}
+\epsilon O_{1}= \frac {1}{T(u)}\left[u^{\mu}h^{\lambda} + \tau^{\mu \lambda}\right]\epsilon_{\lambda} + \epsilon O_{2}.\\[10pt]
\end{equation} 

On the other hand, working out the variation of the $Q^{\mu}(u)$ term arising 
(\ref{FQ1}, \ref{FR}) one 
finds that the majority of the terms yield an 
$\epsilon O_{2}$
contribution except the variation of the 
$R^{\mu}(u)$ terms.
Working out 
$\delta R^{\mu}(u)$
 using formulas (\ref{lema1_02} - \ref{lema3_08}), one finds  an expression identical to that
in (\ref{TQF}) which implies that 
 $Q^{\mu}(u)$ has been chosen
consistently in 
(\ref{TT_04}) and 
(\ref{FR})
and 
this consistency in the choice of the $Q^{\mu}(u)$ plays an important role in proving the frame invariance property of the phenomenological equations.
Under the frame change in (\ref{VRTR}),
a calculation shows
that $S^{\mu}$ in 
 (\ref{TT_05}) 
satisfies the identity:

\begin{equation}\label{TT_06F}
\begin{split}
	S^\mu & = su^{\mu}+\beta(u) h^{\mu}(u)-\Theta(u) n^{\mu}(u)-Q^{\mu}(u) \\
	& = \hat s \hat u^{\mu}+\beta(\hat u) h^{\mu}(\hat u)-\Theta(\hat u) n^{\mu}(\hat u)
-(T^{\mu \lambda}-T_{0}^{\mu \lambda})\delta \beta_{\lambda} \\
&
-Q^{\mu}(\hat u)-\delta R^{\mu}+
\hat \epsilon O_{2}+\hat \epsilon O_{1}.
\end{split}
\end{equation} 

However the term
$-(T^{\mu \lambda}-T_{0}^{\mu \lambda})\delta b_{\lambda}
-\delta R^{\mu}$ cancel out leaving
\begin{equation}\label{TT_05F}
\begin{split}
	S^\mu & = su^{\mu}+\beta(u) h^{\mu}(u)-\Theta(u) n^{\mu}(u)-Q^{\mu}(u) \\
& = \hat s \hat u^{\mu}+\beta(\hat u) h^{\mu}(\hat u)-\Theta(\hat u) n^{\mu}(\hat u)
-Q^{\mu}(\hat u) + 
\hat \epsilon O_{2}+\hat \epsilon O_{1},
\end{split}
\end{equation} 
which establishes the $O_{1}$ invariance of $S^\mu$
under frame change described
by (\ref{VRTR}). This property of
$S^{\mu}$ 
implies that the resulting 
phenomenological equations
are equivalent
under the frame changes described in  (\ref{VRTR}).\\

To get insights into the implications of this invariance property of $S^{\mu}$ shown in (\ref{TT_05F}), let us 
 evaluate the $ Q^\mu(u) $ 
in the energy frame. Since relative to this frame 
 $$ h^\mu(u_E) = 0 ,\quad q^\mu(u_E) = -\dfrac{\rho + P}{n}n^\mu(u_E) $$
using
 (\ref{FQ1}, \ref{FR}),
one finds

\begin{equation}\label{Q_1}
Q^\mu(u_E) = \frac{1}{2}u^\mu_E\left[\beta_0 \pi^2 + \beta_1 n^\nu_E {n_{E}}_{\nu} + \beta_2 \pi^{\lambda \mu}\pi_{\lambda \mu} \right] - \alpha_0 \pi n^\mu_E - \alpha_1 \pi^{\mu \nu}n_{E\nu}
\end{equation} 
where we arrived in this expression by normalizing the $\beta_1$, $\alpha_0$ and $\alpha_1$ coefficient by absorbing terms like $ (\rho + p)n^{-1} $ into their definition. If on the other hand, we 
evaluate the same  term relative to the particle frame where 

$$ q^\mu = h^\mu_N, \quad  n^\mu_{N} = 0 $$
and one introduce new
coefficients 
 $\bar \alpha_{j}$, $j \in (0, 1)$, $\bar \beta_{i}$, 
$i \in (0, 1, 2)$,
one finds

\begin{equation}\label{Q_2}
Q^\mu(u_N) = \frac{u^\mu_N}{2}\left[\bar{\beta}_0\pi^2 + \left(\bar{\beta}_1 + \frac{1}{T(\rho + P)}\right)h^\mu h^\nu + \bar{\beta}_2\pi^{\lambda \nu}\pi_{\lambda \nu}\right] - 
\end{equation}
$$
-\left(\bar{\alpha}_0 - \frac{1}{T(\rho + P)}\right)\pi h^\mu_N - \left(\bar{\alpha}_1 - \frac{1}{T(\rho + P)}\right)\pi^{\mu \nu}h_\nu
$$

A comparison between (\ref{Q_1}, \ref{Q_2}) show  identical structures
provided one defines

\begin{equation}\label{Q_3}
\beta_{0}=\bar \beta_{0},\quad \beta_1 = \bar{\beta}_1 + \frac{1}{T(\rho + P)},\quad  \beta_{2}=\bar \beta_{2},
\quad \alpha_0 = \bar{\alpha}_0 - \frac{1}{T(\rho + P)}, \quad \alpha_1 = \bar{\alpha}_1 - \frac{1}{T(\rho + P)}
\end{equation}
and
these relations have been also derived in Israel \cite{Isr1} and Israel and Stewart 
\cite{Isr2} using different means
that the ones employed in the present work. 
If  formally one introduce $ q^\mu = n^{\mu}_E $ and $ q^\mu = h^\mu_N$ as fields variables
in (\ref{Q_1}) respectively 
(\ref{Q_2}) and use 
(\ref{TT_05F})
to evaluate the phenomenological equations 
in view of 
(\ref{Q_3}),
one obtains identical forms provided terms of $\epsilon O_{1}, O_{2}$ have been neglected,
showing  the $O_{1}$ equivalence of the phenomenological equations relative to the energy or particle frame.
For arbitrary frame changes
generated by $ (u^\mu, \hat u^{\mu})$
in
(\ref{VRTR}), 
the equivalence of the phenomenological equations can also established
but the situation is more complex. For this case, the number of thermodynamical variables increase
and either one can treat $ (q^\mu(u), h^\mu(u) )$ as independent variables or treat $ h^\mu(u) $ and the particle 
drift $ n^\mu(u) $ as independent variables. In either case
(\ref{TT_05F}) shows their equivalence.

We shall leave this section by offering a few comments regarding the causality and stability properties
of the Israel-Stewart transient thermodynamics.\\
 The classical works by Hiscock and Lindblom in \cite{His1},\cite{His3}
and Olson in \cite{Olson}, established 
that linear perturbations  about
a spatially homogeneous equilibrium state\footnote{The identification
of equilibrium states for the Israel-Stewrt theory have been worked out in 
refs. \cite{His1}, \cite{Olson} (see also \cite{Isr1}, \cite{Isr2}).}
are bounded and propagate causally
as long 
as suitable restrictions upon the parameters and the equations of state are imposed.
This conclusion is in sharp contrast to 
 the causality and stability behavior of the Hiscock-Lindblom class of first order theories
 as we have already discussed in sec. (IV).
Gavasinno, Antonelli and Haskell in \cite{GAV6} convincingly argue that
this difference should be traced in thermodynamics
and  it is due to the different structure that the entropy current $S^{\mu}$ has in the two theories
(compare $S^{\mu}$ in (\ref{EFO})
 to that of Israel-Stewart theory shown in  (\ref{TT_05}). It was shown in 
\cite{GAV6} that the total entropy 
based on 
(\ref{EFO})
has no upper bound, while the inclusion of the quadratic 
contributions  in 
  (\ref{TT_05}) for the 
 Israel-Stewart theory 
 implies that the total entropy has an absolute maximum 
presumably representing the equilibrium state.
Finally in a recent work,
Gavassino, Antonelli and Haskell in \cite{GAV1}
 showed for the 
  Israel-Stewart
 transient thermodynamics, perturbations
around global equilibrium states on Minkowski spacetime 
 relative 
to the Eckart (particle) frame  and 
perturbations  relative to
  Landau-Lifshitz (energy) satisfy equivalent equations i.e.
  they are 
 equivalent theories implying that the stability-causality conditions found by Olson
 in \cite{Olson} (in the Landau frame) are just a rewrite of those found by Hiscock and Lindblom 
in  \cite{His1},\cite{His3}  (in the Eckart 
 frame). This is to be contrasted to what occurs for the case of perturbations induced on
 an equilibrium state induced by an Eckart theory or a 
  Landau-Lifshitz theory within the context of the Hiscock-Lindblom class of first order theories\footnote{Our thanks to an 
  anonymous referee who suggested to probe this interconnection between the two classes of theories.}
 that we discussed in section $(IV)$. It would be of interest to examine whether this difference is traced in hydrodynamics or in thermodynamics and this issue is 
  for the moment open. Finally, we mention here the recent results obtained in 
  \cite{BEM} regarding causality, local existence, and uniqueness of solutions of the theory
  within the full nonlinear regime. Overall the theory stands on a very firm ground.

\section{On states satisfying the (LTE) postulate and the Liu-M\"uller-Ruggeri theory}

We finish this paper by turning attention to 
  fluid states compatible with the relativistic (LTE) postulate
and their relation to
 Liu-M\"uller-Ruggeri theory (for an introduction to this theory see \cite{Mul6, Mul4})
and the theory of relativistic fluids of divergence type
developed by Pennisi \cite{Pen}, and independently by Geroch and Lindblom
in \cite{Ger1}. Of these two closely related theories, we shall consider 
only 
the relation between  
states compatible with the relativistic (LTE) postulate
and the
Liu-M\"uller-Ruggeri theory since their relation to 
the theory of fluids of divergence type
requires 
to introduce the generating function and 
the field of Lagrange multipliers and
this needs a more elaborate treatment that will be presented elsewhere \cite{FT2}.\\

We recall that 
within the Liu-M\"uller-Ruggeri theory \cite{Mul6, Mul4},
arbitrary fluid states are described by the $10$ components of the symmetric energy momentum tensor $ T^{\mu \nu} $ and the
$4$ components of the particle  current $ J^\mu $ satisfying\footnote{The structure of eqs.(\ref{RET_1},\ref{RET_2}), the trace free property of $I^{\mu\nu}$ and the relation
 $A^{\mu \nu}{}_{\nu}=m^{2}c^{2}J^{\mu}$ 
 are motivated by relativistic kinetic theory of a simple gas. For an introduction to this theory see refs.
 \cite{Ehl, An1, O3,O4}.}:

\begin{equation}
	\nabla_\mu T^{\mu \nu} = \nabla_\mu J^\mu = 0,
\label{RET_1}	
\end{equation} 
\begin{equation}
	\nabla_\mu A^{\mu \nu \lambda} = I^{\nu \lambda},
\label{RET_2}		
\end{equation}
where $ A^{\mu \nu \lambda} $ is a completely symmetric tensor field 
subject to 
$A^{\mu \nu}{}_{\nu}=m^{2}c^{2}J^{\mu}$
and thus necessarily $\nabla_{\mu}A^{\mu\nu}{}_{\nu}=0$,
while $I^{\mu\nu}$ is symmetric 
and traceless $I^{\mu}{}_{\mu}=0$.
The fields $A^{\mu \nu \lambda}$ and $I^{\mu \nu}$
 are considered to be 
constitutive relations i.e.
$A^{\mu \nu \lambda}=
A^{\mu \nu \lambda}(J^\mu, T^{\mu \nu})$,  $I^{\mu \nu} = I^{\mu \nu}(J^\mu, T^{\mu \nu})$
and this is a special feature of the theory.
In addition to the balance laws
(\ref{RET_1}, \ref{RET_2}), the theory employs an 
 entropy flux vector $ S^\mu $
which is also a constitutive  
function i.e. 
$S^\mu = S^\mu(J^\mu, T^{\mu \nu})$
and
for any solution
$(J^{\mu}, T^{\mu\nu})$
of  (\ref{RET_1}, \ref{RET_2})
obeys:
\begin{equation}
	\nabla_\mu S^\mu(J^{\mu}, T^{\mu\nu}) \geq 0.
\label{RET_entropy1}	
\end{equation}

A major issue  in this theory
is the specification of the
constitutive functions
$A^{\mu \nu \lambda}(J^\mu, T^{\mu \nu})$,  $I^{\mu \nu} = I^{\mu \nu}(J^\mu, T^{\mu \nu})$
and $S^{\mu}=S^{\mu}(J^\mu, T^{\mu \nu})$ as well as the implementation of the
entropy inequality in (\ref{RET_entropy1}).
Liu, M\"uller and Ruggeri in ref.\cite{Mul6} 
(see also ref.\cite{Mul4})
were able 
to construct representations of 
$A^{\mu \nu \lambda}(J^\mu, T^{\mu \nu})$,  $I^{\mu \nu} = I^{\mu \nu}(J^\mu, T^{\mu \nu})$
and $S^\mu = S^\mu(J^\mu, T^{\mu \nu})$
by invoking three principles: the Entropy Principle, the Principle of Relativity,
and the requirement of the Hyperbolic nature of the dynamical equations.
Subsequently,  
by employing the Eckart frame\footnote{The analysis 
 in refs.\cite{Mul6}, \cite{Mul4} has been performed relative to the Eckart frame.
In principle however,
one could 
have chosen an arbitrary velocity field $u^{\mu}$
and 
decompose 
 $J^{\mu}$ and $T^{\mu\nu}$
relative
to this new  $u^{\mu}$
and carry out the analysis 
relative to this new frame.
We are not aware of any 
related work on this issue.}
they worked out predictions of their theory
and compared its predictions of those of transient theory (for details of this comparison see
refs.\cite{Mul6, Mul4}).\\

The purpose in this section is to discuss
states compatible with the relativistic (LTE) postulate
within this theory.\\
For this, let a state 
satisfying the 
(LTE) postulate i.e. it is
described by  $(J^{\mu}, T^{\mu\nu})$ 
obeying
(\ref{RET_1}-\ref{RET_2})	
and subject to the restriction that
the pseudo-angle $\epsilon$ between 
 $(u^{\mu}_{E}, u^{\mu}_{N})$
satisfies
everywhere within the fluid region the condition
$\epsilon<<1$.
Let $u^{\mu}$ be a velocity field 
chosen  in the  region occupied by the fluid
and 
lying within the cone formed by this $\epsilon$,
so that 
local thermodynamical equilibrium prevails.
If
$s(n,\rho)$
denotes the corresponding equilibrium equation of state,
then
$(u^{\mu}, s(\rho, n))$
introduce 
the ''local thermodynamical equilibrium state''
specified by 
$(S^{\mu}_{0}, T^{\mu\nu}_{0}, J^{\mu}_{0})$ (see discussion leading 
to (\ref{33})).
This
$u^{\mu}$
and  the  decompositions
in (\ref{OD},\ref{OD_11}) define
 the fields 

\begin{equation}
(n( u),  \rho(u),  P(u),  h(u)^{\mu},  n(u)^{\mu}, \tau(u)^{\mu\nu})
\label{LMR}	
\end{equation}
and 
these fields describe a state within 
the 
Liu-M\"uller-Ruggeri theory, provided
they satisfy
(\ref{RET_1}, \ref{RET_2}).
However, in order to impose these equations, we need a representation of 
$A^{\mu \nu \lambda}(J^\mu, T^{\mu \nu})$,  $I^{\mu \nu} = I^{\mu \nu}(J^\mu, T^{\mu \nu})$
and their specification is a delicate problem.
Since in this work, 
we are interested only  in the behavior of small deviations 
away from the local ''thermodynamical equilibrium state''
specified by
$(u^{\mu}, s(\rho, n))$
(or equivalently by
$(S^{\mu}_{0}, T^{\mu\nu}_{0}, J^{\mu}_{0})$), we
 set
\begin{equation}
 T^{\mu\nu}= T^{\mu\nu}_{0}+\delta T^{\mu\nu},\quad
J^{\mu}=J^{\mu}_{0}+\delta J^{\mu}.
\label{44a}
\end{equation}
and  take advantage of the 
constitutive nature of
$A^{\mu \nu \lambda}(J^\mu, T^{\mu \nu})$,  $I^{\mu \nu} = I^{\mu \nu}(J^\mu, T^{\mu \nu}).$
This later property allows us to write:
\begin{equation}
\begin{split}
A^{\mu \nu \lambda}(T^{\mu \nu}, J^{\mu}) & =
A^{\mu \nu \lambda}( T^{\mu \nu}_{0}+
\delta T^{\mu\nu}, J^{\mu}_{0}+\delta J^{\mu}) \\
	& = A^{\mu \nu \lambda}( T^{\mu \nu}_{0},J^{\mu}_{0})+
\frac {\delta A^{\mu \nu \lambda}}{\delta T{} ^ {\alpha\beta}}\delta T^{\alpha\beta} \\
& +\frac {\delta A^{\mu \nu \lambda}}{\delta J{} ^ {\alpha}}\delta J^{\alpha}+ O(\delta J^{\mu})^2+
O(\delta T^{\mu\nu})^2,
\end{split}
\end{equation}
\begin{equation}
\begin{split}
I^{ \nu \lambda}(T^{\mu \nu}, J^{\mu}) & = 
I^{\nu \lambda}( T^{\mu \nu}_{0}+
\delta T^{\mu\nu}, J^{\mu}_{0}+\delta J^{\mu}) \\
	& = I^{ \nu \lambda}( T^{\mu \nu}_{0},J^{\mu}_{0})+
\frac {\delta I^{\nu \lambda}}{\delta T{} ^ {\alpha\beta}}\delta T^{\alpha\beta}
\\
& + \frac {\delta I^{ \nu \lambda}}{\delta J{} ^ {\alpha}}\delta J^{\alpha}+ O(\delta J^{\mu})^2+
O(\delta T^{\mu\nu})^2,
\end{split}
\end{equation}

where in these expansions
we kept terms describing only first order
deviations
from the state of the local equilibrium field.\\
The symmetries of
$A^{\mu \nu \lambda}$ and the condition
 $A^{\mu \nu}{}_{\nu}=m^{2}c^{2}J^{\mu}$ 
allows us to propose
a representation of 
 $A^{\mu \nu \lambda}( T^{\mu \nu}_{0},J^{\mu}_{0}):=A^{\mu \nu \lambda}_{0}(u)$ 
 involving only  terms that appear in the representation of the local equilibrium field
 $(S^{\mu}_{0}, T^{\mu\nu}_{0}, J^{\mu}_{0})$.
These
symmetries and the  trace condition dictate that 
$A^{\mu \nu \lambda}_{0}(u)$
should be of the form\footnote{For this section indices enclosed in a parenthesis indicate symmetrization, thus
$
u^{( \mu}u^{\nu}u^{\lambda)}=\frac {1}{6}( u^\mu u^{\nu} u^{\lambda}+
 u^\nu u^{\lambda} u^{\mu} +u^\lambda u^{\mu} u^{\nu})$.}

\begin{equation}
A^{\mu \nu \lambda}_{0}(u)=
2 {\gamma_{1}}u^{( \mu}u^{\nu}u^{\lambda)} + (nm^{2}+\gamma_{1})g^{(\nu\lambda}u^{\mu)},
\label{44c}
\end{equation}
where 
$\gamma_{1} $ is 
a function of $(n,\rho)$ to be determined and $ m $ is  a mass scale\footnote{The right hand-side of
(\ref{44c}) as well as of 
(\ref{45d}), are indeed functions of $T^{\mu\nu}$ and $J^{\mu}$. One can see that 
by expressing the fields
in
(\ref{LMR})  in terms of $T^{\mu\nu}$ and $J^{\mu}$.}.
Similarly the symmetry requirements  
upon $I^{\mu\nu}$
permit us to
 propose 
that the leading contributions of the
$I^{\nu \lambda}(T^{\mu \nu}, J^{\mu})$
 to be vanishing i.e.
$I^{\nu \lambda}( T^{\mu \nu}_{0}, J^{\mu}_{0}):=0$
and thus to set the expansion of $I^{\nu \lambda}$ in
(\ref{45d}) to have the form

\begin{equation}\label{45d}
\begin{split}
I^{ \nu \lambda}(T^{\mu \nu}, J^{\mu}) & = \frac {\delta I^{\nu \lambda}}{\delta T{} ^ {\alpha\beta}}\delta T^{\alpha\beta}
+\frac {\delta I^{ \nu \lambda}}{\delta J{} ^ {\alpha}}\delta J^{\alpha}\\
& = \delta_{1}\pi (g^{\mu\nu}+4u^{\mu}u^{\nu}) 
+\delta_{2}\pi^{\mu\nu}\\
& 
+\delta_{3}(q^{\mu}u^{\nu}+q^{\nu}u^{\mu}),
\end{split}
\end{equation}
where
$q^{\mu}$ is the invariant heat flux defined in 
(\ref{lema2_01})
and  the coefficients 
$\delta_{i}, i\in (1,2,3)$
 are unknown functions of $(n,\rho)$.
With these choices, the fields in
(\ref{LMR}) satisfy

\begin{equation}
\begin{split}
\nabla_{\mu}T^{\mu\nu} & = \nabla_{\mu}[\rho(u)u^{\mu}u^{\nu}+P(u)\Delta(u)^{\mu\nu}+h(u)^{\mu}u^{\nu} \\ & +h(u)^{\nu}u^{\mu}+\tau(u)^{\mu\nu}]=0,
\end{split}
\label{LMR1}
\end{equation}
\begin{equation}
	\nabla_{\mu} J^{\mu}=\nabla_{\mu}[n(u)u^{\mu}+n(u)^{\mu}]=0
\label{LMR2}	
\end{equation}
\begin{equation}
\begin{split}
	\nabla_{\mu} A^{\mu \nu \lambda}_{0}(u) & =	
\delta_{1}\pi(g^{\mu\nu}+4u^{\mu}u^{\nu})
+\delta_{2}\pi^{\mu\nu} \\
&
+\delta_{3}(q^{\mu}u^{\nu}+q^{\nu}u^{\mu}).
\end{split}	
\label{LMR3}	
\end{equation}

We leave aside for the moment 
the implementation of the entropy law in (\ref{RET_entropy1}) since as we have already mentioned, its implementation requires
to introduce the generating function,
the field of Lagrange multipliers 
and apply of Liu's procedure \cite{Liu}
for the implementation of 
of the second law.
Since these issue 
need altogether a different machinery
they will be discussed in a separate account in connection 
to  fluids of divergence type \cite{FT2}. For the rest of this section we shall treat
(\ref{LMR1}-\ref{LMR3}) as the equations that  fluid states
compatible to the relativistic (LTE) postulate are required to satisfy.\\

Let us suppose now that 
the system (\ref{LMR1}-\ref{LMR3})
admits a solution 
$(u^{\mu}, n( u),  \rho(u),  P(u),  h(u)^{\mu},  n(u)^{\mu}, \tau(u)^{\mu\nu})$
and let us
 consider a 
frame change
described by 
 \begin{equation}
u^{\mu}\to \hat u^{\mu}=u^{\mu}+\hat \epsilon^{\mu},\quad \hat\epsilon \leq \epsilon,
\label{FRCH1}	
\end{equation}
where $\hat u^{\mu}$ 
has been chosen to
 lie
within the cone of the opening pseudo-angle 
$\epsilon<<1$
and 
relative to this new frame, we define 
fields $(\rho(\hat u), n(\hat u), P(\hat u), h^{\mu}(\hat u), \tau^{\mu\nu}(\hat u), n^{\mu}(\hat u))$
by applying the same procedure as 
for the case of first order theory and the case of transient theory.
 It is easily  seen that
the transformed fields 
 $(\hat u^{\mu}, \rho(\hat u), n(\hat u), P(\hat u), h^{\mu}(\hat u), \tau^{\mu\nu}(\hat u), n^{\mu}(\hat u))$ 
satisfy to an $O_{1}$ accuracy
the conservation laws in
(\ref{RET_1}). However matters complicate in proving that 
the transformed fields satisfy
  to an $O_{1}$ accuracy
 equation 
  (\ref{LMR1}-\ref{LMR3}).
 To facilitate 
matters, we note that (\ref{LMR3}) imply

\begin{equation}\label{Sec:LMR1}
u_\nu u_\lambda \nabla_\mu A^{\mu \nu \lambda}_{0}(u) = 3\delta_1 \pi(u),
\end{equation}
\begin{equation}\label{Sec:LMR2}
\Delta_{\rho \nu}(u)u_\lambda \nabla_\mu A^{\mu \nu \lambda}_{0}(u) = \delta_3 q_\rho(u),
\end{equation}
\begin{equation}\label{Sec:LMR3}
\left(\Delta_{\alpha\nu}(u) \Delta_{\beta \lambda}(u) - \frac{1}{3}\Delta_{\alpha \beta}(u)\Delta_{\nu \lambda}(u)\right)\nabla_\mu A^{\mu \nu \lambda}_{0}(u) = \delta_2 \pi_{\alpha \beta}(u),
\end{equation}
and the idea is to show that
these equations remain form $O_{1}$ invariant under a change of the rest frame 
described by (\ref{LMR3}). 
Based on the 
representation of
$A^{\mu \nu \lambda}_{0}(u)$ in 
(\ref{44c}), then a calculus shows that
that under the frame change 
in (\ref{FRCH1})  the following formula holds:

\begin{equation}
\nabla_{\mu} A^{\mu \nu \lambda}_{0}(u)=\nabla_{\mu} A^{\mu \nu \lambda}_{0}(\hat u)-\nabla_{\mu}[2 \gamma_{1}
	\hat \epsilon^{( \mu}\hat u^{\nu}\hat u^{\lambda)} +	
(nm^{2}+\gamma_{1})g^{(\nu\lambda} \hat \epsilon ^{\mu)} +\hat \epsilon O_{1}].
\label{LMR33}	
\end{equation} 
Using this formula, 
it is straightforward to verify that (\ref{Sec:LMR1}, \ref{Sec:LMR2}, \ref{Sec:LMR3}) remains form invariant to $O_1$ deviation from the state of local equilibrium taking into account that $ \nabla_\mu u^\mu = O(1) $, $ \nabla_\mu \epsilon^\mu = O(1) $ and always $ u^\mu \epsilon_\mu = 0 $.\\
In summary 
within the class
of the Liu-M\"uller-Ruggeri theory, states satisfying the relativistic (LTE) postulate
satisfy eqs that remain invariant 
under a frame change described 
in (\ref{FRCH}) 
and thus exhibit the same behavior as 
states near equilibrium for the first order theories and the transient thermodynamics.
We shall discuss further properties of states satisfying the relativistic (LTE) postulate in light of some recent
 advances on the theory of divergence type fluids reported lately in refs. \cite{GAV1} (see also \cite{GAV7}) .

\section{Discussion}

In this article, we have introduced a special class 
of fluid states satisfying (or been compatible to) 
the relativistic (LTE) postulate.
Our motivation to introduce this class has been partially induced by 
the role of the (LTE)
postulate within the Onsager-Eckart theory
and largely by Israel's ideas on the description of relativistic fluid states
that are considered to be near equilibrium within the transient thermodynamics.\\
We started the article by pointing out
the implications of the (LTE) postulate within the Onsager-Eckart theory
and stressed that states compatible
with the (LTE) postulate are special. Firstly they permit  to introduce local thermodynamical 
variables and secondly
their physical entropy is determined by these
local variables\footnote{Because of these properties, often in the literature and even within the context of
relativistic fluids, one encounter the term ''theories based on the local equilibrium hypothesis'' . With this term 
one understands
theories whose independent variables are those that determine equilibrium states. Clearly this definition is not wide enough, for instance transient thermodynamics, incorporates fluxes as independent variables and these variables
do not enter in the description of equilibrium states.}.
The 
extension of the classical  (LTE) postulate to the relativistic regime
proposed in this paper, define
fluid states compatible with the 
relativistic (LTE) postulate
 as states subject to two restrictions.\\
The first one
requires that the state should allow 
the 
attachment of 
a ''local thermodynamical equilibrium'' state
 and 
this
attachment
requires 
that the four velocity $u^{\mu}$ 
ought to be chosen so that 
this 
 $u$-observer,
relative to her/his local rest frame, 
detects a collision time scale which is much shorter 
than any other macroscopically defined time scale.
Under this condition,
an "equilibrium equation of state'' of the form
$s=s(\rho, n)$ may be postulated to exist
and thus 
''thermal'' aspects such as local temperature and local chemical or thermal potential
appear\footnote{Notice, that even though any observer with an arbitrary
four velocity $u^{\mu}$, using 
the state variables  $(J^{\mu}, T^{\mu\nu})$
can always define along his/her world line particle density $n$ energy density $\rho$ energy flux $h^{\mu}$ y stresses $\tau^{\mu\nu}$,
still  ''thermal'' aspects do not appear unless an 
"equilibrium equation of state'' 
$s=s(\rho, n)$ is postulated to exist.}.
The second restriction requires the state to be 
a state near equilibrium in Israel's  sense, meaning that everywhere
within the region occupied by the fluid, the ''angle $\epsilon$'' formed by
the four velocity that defines the energy frame and the four velocity that defines 
 the particle frame, satisfies $\epsilon<<1$.
Whenever any
$u^{\mu}$
lies within the cone of opening angle $\epsilon$,
then the appearance of the 
class of admissible frames 
and the treatment of
deviations of the physical state, from the state
of 
''local thermodynamical equilibrium'' 
follows the root that we presented in 
sections $(III, VI)$.\\
Parenthetically we
add, 
that in order that
 a fluid 
state be compatible with
the relativistic (LTE) postulate,
it is necessary  that it satisfies simultaneously both of the above mentioned restrictions. 
One can imagine fluid states where either the 
attachment  of 
a ''local thermodynamical equilibrium'' 
it is not possible
or the state it is not a state near equilibrium in Israel's sense. In other case, one does not get the benefits arising from
the existence of admissible frames.

Although in this paper, we 
examined properties of states satisfying the relativistic  (LTE) postulate 
within the context of the Hiscock-Lindblom class of first order theories, transient thermodynamics, the (BDNK)
theory, and the  
Liu-M\"uller-Ruggeri theory,
it would be of interest to include in this list, the
class of relativistic fluids of divergence type
in view on some recent advances in this theory
reported by Gavassino, Antonelli and Haskell 
in refs \cite{GAV1}, \cite{GAV7}.\\
Overall the analysis in this paper shows that that
within the Hiscock-Lindblom class of first order theories, 
states satisfying the relativistic  (LTE)-postulate 
they do not seem to offer any advantage.
Like arbitrary states within this theory, 
they violate causality and equilibrium states 
are unstable. However, maters are different 
within the (BDNK) theory
and transient thermodynamics. As far as the latter theory is concerned, we have argued in section
$V$, that the theory deals exclusively with states 
obeying 
the relativistic (LTE)-postulate and in reaching  this conclusion, 
 refs \cite{His4}, \cite{GAV7}, \cite{BEM} are very relevant.
 As far as the
 former theory is concerned,
 in section $(IV)$,
 we presented evidences  
 suggesting  that
states obeying 
the relativistic  (LTE)-postulate are states 
referred as states near equilibrium within the
(BDNK) formalism.
In fact we believe, that all states 
within the (BDNK) theory where  
the gradient series expansion is meaningful,
are in fact states compatible to  
the relativistic (LTE)-postulate, although at this point this is a conjectural claim. In that regard,
it would be of interest, starting from the relativistic Boltzmann equation
and via suitable expansion investigate whether one may generate fluid states where the pseudo-angle $\epsilon$
formed by $u_{E}^{\mu}$ and 
$u_{N}^{\mu}$ satisfies
 everywhere the condition$\epsilon<<1$. 
 So far 
 the relativistic Boltzmann equation
  has offered considerable insights in the derivation of  
relativistic hydrodynamic equation (see for instance \cite{Isr2}, \cite{Nor1}, \cite{Den} and references therein)
and may be further efforts 
one gets insights into
the nature of the geometry of possible rest frames may be helpful.
Currently we have that issue under active investigation.\\
We finish this paper
by mentioning that
in the literature
there have been advanced alternative 
definitions of states obeying the relativistic (LTE) postulate
and 
these definitions either
originate 
from the quantum statistical framework 
(see for instance \cite{Bec} and reference therein)
It would be of some interest to compare these definitions
and their predictions to the one proposed in this work.

  \section{Acknowledgments}\label{ACK}

Our warm thanks to the members of the relativity group at the IFM Univ. Michoacana
for multiple discussions. Special thanks to O. Sarbach whose continuing poking
 lead to the development of this work.
The research of T.Z was supported  by a Grant from CIC-UMSNH.
During the completion of this work J.F.S was funded by 
CONACYT by a post doctoral fellowship
through the project A1 -S-31269.

\end{document}